\documentclass[a4paper,11pt]{article}
\pdfoutput=1
\usepackage{jcappub}
\usepackage[T1]{fontenc}
\usepackage{algorithm}
\usepackage{algpseudocode}
\usepackage{siunitx}

\DeclareSIUnit\parsec{pc}
\DeclareSIUnit\h{h}
\DeclareSIUnit\msolar{\ensuremath{M_\odot}}

\author[a]{{Sebastian~Seehars,}}
\author[a,b]{{Aseem~Paranjape,}}
\author[a,c]{{Amadeus~Witzemann,}}
\author[a]{{Alexandre~Refregier,}}
\author[a]{{Adam~Amara,}}
\author[a]{and {Joel~Akeret}}

\affiliation[a]{ETH Zurich, Institute for Astronomy, Department of Physics, Wolfgang Pauli Strasse 27, 8093 Zurich, Switzerland}
\affiliation[b]{Inter-University Centre for Astronomy and Astrophysics, Ganeshkhind, Post Bag 4, Pune 411007, India}
\affiliation[c]{Astrophysics, Cosmology and Gravity Centre, Department of Mathematics and Applied Mathematics, University of Cape Town, Rondebosch 7701, South Africa}

\emailAdd{sebastian.seehars@phys.ethz.ch}
\emailAdd{aseem@iucaa.in}
\emailAdd{wtzama001@myuct.ac.za}
\emailAdd{alexandre.refregier@phys.ethz.ch}
\emailAdd{adam.amara@phys.ethz.ch}
\emailAdd{joel.akeret@phys.ethz.ch}

\title{Simulating the Large-Scale Structure of HI Intensity Maps}

\abstract{
Intensity mapping of neutral hydrogen (HI) is a promising observational probe of cosmology and large-scale structure. We present wide field simulations of HI intensity maps based on N-body simulations of a $2.6\, {\rm Gpc / h}$  box with $2048^3$ particles (particle mass $1.6 \times 10^{11}\, {\rm M_\odot / h}$). Using a conditional mass function to populate the simulated dark matter density field with halos below the mass resolution of the simulation ($10^{8}\, {\rm M_\odot / h} < M_{\rm halo} < 10^{13}\, {\rm M_\odot / h}$), we assign HI to those halos according to a phenomenological halo to HI mass relation. The simulations span a redshift range of $0.35 \lesssim z \lesssim 0.9$ in redshift bins of width $\Delta z \approx 0.05$ and cover a quarter of the sky at an angular resolution of about $7'$. We use the simulated intensity maps to study the impact of non-linear effects and redshift space distortions on the angular clustering of HI. Focusing on the autocorrelations of the maps, we apply and compare several estimators for the angular power spectrum and its covariance. We verify that these estimators agree with analytic predictions on large scales and study the validity of approximations based on Gaussian random fields, particularly in the context of the covariance. We discuss how our results and the simulated maps can be useful for planning and interpreting future HI intensity mapping surveys.
}

\arxivnumber{1509.01589}

\notoc
\begin{document}
\maketitle
\flushbottom

\section{Introduction}
\label{sec:introduction}

Intensity mapping of atomic and molecular line transitions is emerging as a promising probe of cosmological evolution as well as of the connection between galaxy evolution and the growth of large-scale structure in the Universe \cite{2015aska.confE..19S,Croft:2015tm,Chang:2010ds,Masui:2013jd,Switzer:2013ea,Comaschi:2015un, Silva:2015dd, Gong:2011gw, Li:2015uc, Mashian:2015ub}. The ability to detect a line over wide areas of the sky implies that one has accurate redshift information over large cosmological volumes and forecasts indicate great potential for the recovery of cosmological information from intensity mapping surveys~\cite{Wyithe:2007cm,2008PhRvL.100p1301L,2009JCAP...10..030V,2015ApJ...803...21B,2015aska.confE...4C}. The \SI{21}{cm} line of neutral hydrogen (HI) is a particularly promising future probe, being the target of several dedicated efforts using radio telescopes such as BAOBAB~\cite{2013AJ....145...65P}, BAORadio~\cite{2012CRPhy..13...46A}, BINGO~\cite{Battye:2013kh}, and CHIME~\cite{Bandura:2014fx}. A successful HI intensity mapping experiment will map the integrated emission of HI within the beam of the instrument over large parts of the sky. Due to the high frequency resolution of radio telescopes, such experiments would produce maps of the large-scale structure of HI in the Universe within thin redshift slices. The angular and redshift resolution in these surveys will correspond to comparable physical scales.

Predicting the constraining power of \SI{21}{cm} experiments requires accurate knowledge of the expected signal. Several approaches have been developed for this purpose. Analytic methods based on perturbation theory and the halo model can be used to predict some of the statistical properties of the signal, such as the power spectrum~\cite{Wyithe:2007cm,Battye:2013kh,2014PhRvD..89h3010P,2015ApJ...803...21B,Padmanabhan:2015vg}. For a fast production of wide field intensity maps, simulations have been developed based on log-normal random fields~\cite{Alonso:2014bq}. A number of different numerical approaches have also been used to understand the effect of non-linear and baryonic processes on the signal. They range from hydrodynamical simulations~\cite{Duffy:2012dl, Dave:2013bf} that model both collisionless dark matter and the hydrodynamics of baryons to approaches where the HI is simply assigned to resolved halos from dark matter only simulations~\cite{Bagla:2010fu, Khandai:2011is}.

In this paper, we follow an alternative and complementary approach by simulating low-redshift HI intensity maps ($z<1$) using a combination of dark matter fields from N-body simulations and a halo model prescription for assigning HI to sub-resolution dark matter halos. The advantage of this approach is that it produces maps that have both realistic clustering from N-body simulations and large volumes, as necessary for future HI intensity mapping experiments. For this purpose, we use N-body simulations of a \SI{2.6}{\per\h\giga\parsec} box~\cite{Wechsler15} with a mass resolution of $\SI{1.6d11}{\per\h\msolar}$. Below this mass resolution, we model the halo distribution using a conditional halo mass function~\cite{2002MNRAS.329...61S}. We then generate HI intensity maps by assigning HI mass to the halos following a phenomenological prescription~\cite{Padmanabhan:2015vg}.

Using the maps, we study the wide field clustering properties of the HI intensity fluctuations, focusing in particular on the impact of non-linearities. HI intensity maps are characterized by having both the wide field and the relatively low angular resolution of Cosmic Microwave Background (CMB) maps and the three dimensional, non-Gaussian large-scale structure of galaxy surveys. We therefore measure the angular power spectrum $C_\ell$ of the HI intensity maps using the pseudo $C_\ell$ estimator~\cite{1973ApJ...185..413P} and the publicly available PolSpice estimator~\citep{2001ApJ...561L..11S, Chon:2004ki}. These estimators have been developed for the CMB and have also been applied to galaxy surveys (see e.g.~\cite{Thomas11042011,2015arXiv150705598B}). We assess their relative performance in this new regime and compare the results to expectations from analytic models.

We also study the covariance of the angular power spectrum estimators which is of particular importance for interpreting upcoming HI surveys and for assessing their constraining power. As for the power spectrum, we apply covariance estimators used for the CMB and galaxy surveys. We estimate the covariance of the aforementioned $C_\ell$ estimators using jackknife and analytical estimates based on Gaussian statistics and compare the results to the covariance estimated from a suite of simulated HI intensity maps.

The paper is organized as follows. In section \ref{sec:intensity_mapping} we briefly review the principles of intensity mapping. The details of our simulations are discussed in section \ref{sec:simulating_h_i_intensity_maps}. In section \ref{sec:the_angular_power_spectrum}, we use these simulations to study their angular power spectra. The covariance of the estimated angular power spectra is analyzed in section \ref{sec:covariance_of_angular_power_spectra}. We summarize and conclude in section \ref{sec:conclusions}.

\section{HI intensity mapping}
\label{sec:intensity_mapping}

We first briefly review the principles of HI intensity mapping at low redshifts. We discuss the observational principles for the example of a single dish experiment before giving a summary of the theoretical expectations for the clustering of the large-scale HI distribution. For more details see e.g.~\cite{Pritchard:2012ja} for a review.

\subsection{Cosmological 21 cm line emission}
\label{sub:intensity_mapping}

Mapping the matter distribution in our Universe is a central part of observational cosmology. Traditionally, this is done by studying the distribution and the shapes of galaxies in wide field surveys. In HI intensity mapping, the idea is to study the distribution of HI by mapping the distribution of redshifted flux from its \SI{21}{cm} line emission. After reionization, the bulk of HI is expected to be clumpy and associated with galaxies. E.g., at redshifts $z\lesssim1.5$,  damped Lyman-$\alpha$ (DLA) systems have been found associated with MgII and FeII absorption systems \cite{2000ApJS..130....1R,2006ApJ...636..610R}. HI emission line stacking techniques applied to star forming galaxies at low redshifts $z\sim0.2$-$0.3$ have yielded measurements of $\Omega_{\rm HI} \sim 10^{-3}$ \cite{Lah:2007fo,Lah:2009hf}. And finally, HI emission studies in the local Universe ($z < 0.03$) have yielded individual detections of a large number of HI-rich galaxies, with $\Omega_{\rm HI} \sim {\rm few\,}\times10^{-4}$ \cite{Martin:2010du,2005MNRAS.359L..30Z}. See also \cite{Padmanabhan:2014ug} for a compilation of HI observations. In this work, we will distribute HI in dark matter halos using a phenomenological prescription proposed in \cite{Padmanabhan:2015vg} that is consistent with these observations of $\Omega_{\rm HI}$ and the hypothesis that HI is typically associated with star forming galaxies\footnote{We do note that recent measurements of the clustering of DLA systems at redshifts $2\lesssim z\lesssim3$ \cite{2012JCAP...11..059F} are discrepant with direct HI observations on which our model is built \cite{Padmanabhan:2015vg}. The resolution of this discrepancy is unclear at present, and we will return to this issue in future work.}.

When mapping the flux from individual clouds of neutral hydrogen for example with the large beam of a single dish radio telescope, the flux of discrete but non-resolved sources on the sky is integrated within a physical volume defined by the size of the beam and a range in frequency. Very large single dish instruments like the Green Bank Telescope with a diameter of 100 m have a resolution of order \SI{10}{\arcminute} at $z \approx 1$. In frequency, however, radio telescopes can have channels with a bandwidth of $\sim 1$ MHz and smaller, corresponding to redshift bins of width $\Delta z \approx 0.003$ at redshift $z \approx 1$. In co-moving distances, the intensity maps from such an instrument would hence have comparable angular and radial resolution of order $\sim \SI{10}{\per\h\mega\parsec}$ at this redshift. In the more distant future, however, intensity mapping surveys with long baseline interferometers like the SKA will be able to improve the angular resolution by many orders of magnitude.

Given an average density of neutral hydrogen $\rho_{\rm HI}(z)$ at redshift $z$ or equivalently its ratio to the critical density today $\Omega_{\rm HI}(z)$, the average brightness temperature of \SI{21}{cm} flux is given by~\cite{Battye:2013kh}
\begin{equation}
	\bar T(z) \simeq \SI{44}{\micro\kelvin} \left(\frac{\Omega_{\rm HI}(z)h}{2.45\times10^{-4}} \right)\frac{(1+z)^2}{E(z)}
	\label{eq:tbar}
\end{equation}
with $E(z)=H(z)/H_0$ being the normalized Hubble parameter. At redshift $z \approx 1$, this corresponds to brightness temperatures of approximately \SI{100}{\micro\kelvin} and is therefore roughly four orders of magnitude sub-dominant to the brightness of our own galaxy at these frequencies which is at the $\sim \SI{1}{\kelvin}$ level even at high galactic latitudes. One of the key challenges of this technique is hence the ability to separate the signal from the Galaxy and other extra-Galactic radio sources~\cite{Ansari:2012hy, Wolz:2014hb, Alonso:2014dz, BigotSazy:2015ch}.

\subsection{Large-scale structure of HI}
\label{sub:large_scale_structure_of_hi}

The fluctuations of \SI{21}{cm} brightness temperatures are expected to be a biased version of the fluctuations in the matter density field. Hence, the Fourier transformed temperature fluctuations in the map at large scales can be written as
\begin{equation}
	\delta T(k, z) = \bar T(z) b_{\rm HI}(k, z) \delta(k, z),
	\label{eq:tkz}
\end{equation}
where $\delta$ is the dark matter overdensity and $b_{\rm HI}(k, z)$ is the bias of HI relative to $\delta$.

On even larger scales, the power spectrum of the matter density fluctuations is well described by linear theory predictions and the halo bias is scale independent. Assuming a simple relation $m_{\rm HI}(m)$ between HI mass $m_{\rm HI}$ and halo mass $m$, the large scale bias of HI $b_{\rm HI}(z)$ can be predicted from the halo mass function using:
\begin{equation}
	b_{\rm HI}(z) = \int dm \,\mathcal N(m,z) \,\frac{m_{\rm HI}(m)}{\bar\rho_{\rm HI}(z)}\,b(m,z)\,,
	\label{eq:bHI-halomodel}
\end{equation}
where $\bar\rho_{\rm HI}(z) = \int dm \,\mathcal N(m,z) \,m_{\rm HI}(m)$ is the mean density of HI, $\mathcal N(m,z)$ is the unconditional differential number density of halos and $b(m,z)$ is the corresponding halo bias (we use the Sheth-Tormen forms \cite{2002MNRAS.329...61S}; see also Section~\ref{sub:distributing_halos}). The large scale power spectrum of HI is hence expected to be well described by
\begin{equation}
	P_{\rm HI}(k, z) \simeq \left[ \bar T(z) b_{\rm HI}(z) D(z) \right]^2 P(k),
	\label{eq:hipk}
\end{equation}
where $P(k)$ is the linear theory power spectrum of the matter overdensities at $z = 0$ and $D(z)$ is the linear growth factor.

As wide-field surveys probe a spherical region rather than a cartesian box, different methods for measuring the two-point correlations have been proposed and compared (see e.g.~\cite{Asorey:2012fx, Nicola:2014jy, Lanusse:2015dn}). One approach is to perform a tomographic analysis based on the angular correlations (or equivalently the power spectrum) of the field within bins of redshift. This approach is convenient for intensity mapping as techniques for its estimation from maps are well established from the CMB~\cite{1997PhRvD..55.5895T, 1998PhRvD..57.2117B, Szapudi:2000vt, 2001ApJ...561L..11S, 2002ApJ...567....2H, Efstathiou:2004bh, Chon:2004ki}, there is no need for assuming a cosmology during the analysis~\cite{2011PhRvD..84f3505B,2011PhRvD..84d3516C}, and cross-correlating different surveys is straight forward~\cite{Eriksen:2015ex}. Furthermore, intensity mapping surveys will produce maps of brightness temperature fluctuations within bins in frequency and thus a tomographic analysis within these bins does not erase any information. The angular power spectrum $C_\ell(z,z')$ of the temperature fluctuations is related to $P(k)$ via
\begin{equation}
	\begin{aligned}
			C_\ell(z,z') = \frac {2} {\pi} &\int d\tilde z\, W(\tilde z)\bar T(\tilde z) D(\tilde z) b_{\rm HI}(\tilde z) \int d\tilde z'\, W'(\tilde z') \bar T(\tilde z') D(\tilde z') b_{\rm HI}(\tilde z')\\
			& \times \int k^2 dk\, P(k) j_\ell(kR(\tilde z)) j_\ell(kR(\tilde z')),
	\end{aligned}
	\label{eq:lintheory}
\end{equation}
where $W, W'$ are the redshift window functions for the two tomographic bins around $z, z'$ and $R(z)$ is the co-moving distance to redshift $z$.

An additional contribution to the angular power spectrum arises from the shot noise of the discrete sources. Given a population of unresolved, Poisson distributed point sources with neutral hydrogen mass $m_{\rm HI}$ distributed over the full sky, the power spectrum of the resulting intensity map is given by $C_\ell^{sn} = \left(\frac {\bar T} {\bar m_{\rm HI}}\right)^2 \int dm_{\rm HI}\, n_{\rm HI}(m_{\rm HI})\, m_{\rm HI}^2$ with $n_{\rm HI}(m_{\rm HI})$ being the differential source count per steradian and $\bar m_{\rm HI}(z)$ being the mean hydrogen mass of the population~\cite{1996MNRAS.281.1297T}. For our maps, this contribution will however turn out to be negligible ($<\SI{5d-3}{\micro\kelvin}^2$ at the lowest redshift) due to the large number of low mass halos.

\section{Simulations of HI intensity maps}
\label{sec:simulating_h_i_intensity_maps}

In the following, we describe how we construct simulated HI intensity maps using the dark matter field of N-body simulations, the halo model, and a phenomenological prescription for assigning HI mass to halos.

\subsection{Matter density fields}
\label{sub:simulating_density_fields}

We use \num{10} N-body simulations run with the L-Gadget2 code (based on Gadget2~\cite{2001NewA....6...79S,2005MNRAS.364.1105S}). These simulations are also being used to simulate galaxy catalogs for the Dark Energy Survey~\cite{Wechsler15, 2013AAS...22134107B, 2013AAS...22135221B}, and a subset of these catalogs have been used previously in several Dark Energy Survey studies~\cite{2015PhRvD..92b2006V,2015arXiv150705909B,2015arXiv150705647L,2015arXiv150705598B,2015arXiv150705353P}. In this work, we use only the dark matter distribution from these simulations. Each of the \num{10} simulations contains $\num{2048}^3$ particles of mass $\SI{1.6d11}{\per\h\msolar}$ in a \SI{2.6}{\per\h\giga\parsec} box and produces lightcones on the fly by writing all particles that cross the lightcone surface at each time step of the simulation. The simulations use a flat $\Lambda$CDM cosmology with matter density relative to critical $\Omega_m = 0.286$, baryon density relative to critical $\Omega_b = 0.047$, Hubble parameter $h = 0.72$, root mean squared density fluctuations at $8 h^{-1}$Mpc given by $\sigma_8 = 0.82$ and spectral index of the primordial power spectrum $n_s = 0.96$. We use the lightcone output to create HEALPix\footnote{\url{http://healpix.sourceforge.net}}~\cite{2005ApJ...622..759G} maps of the projected matter distribution with an angular resolution of about \SI{7}{\arcminute} ($n_{\rm side} = 512$) by counting the number of particles per pixel, where a pixel is the cosmological volume defined by a HEALPix cell and a redshift bin. We choose the redshift bins such that they correspond to the redshift of the \SI{21}{cm} line in equally spaced bins in frequency between \SI{750}{\mega\hertz} and \SI{1050}{\mega\hertz} with a bandwidth of \SI{25}{\mega\hertz} per channel. This results in \num{12} maps of the radially averaged density field between redshifts $z \simeq 0.35$ and $z \simeq 0.89$ with negligible shot noise and redshift bin width from $\Delta z \simeq 0.033$ at the lowest to $\Delta z \simeq 0.061$ at the highest redshift. In co-moving scales, our pixels have an angular and radial size of \SI{2}{\per\h\mega\parsec} and \SI{82}{\per\h\mega\parsec} at $z \simeq 0.369$ and \SI{4}{\per\h\mega\parsec} and \SI{110}{\per\h\mega\parsec} at $z\simeq 0.863$, respectively. As the co-moving distance to $z = 0.89$ (\SI{2.1}{\per\h\giga\parsec}) is bigger than half the size of the box (\SI{1.3}{\per\h\giga\parsec}), a full-sky lightcone would overlap in the simulated box. To balance overlap and sky coverage, we analyze four quadrants of the lightcone separately. This procedure hence yields \num{40} realizations of the matter density field, each covering a quarter sky. The volume of one quadrant corresponds to roughly $53\%$ of the total box volume. $11\%$ of the quadrant volume is overlapping in the box, but most of this overlapping volume appears at different redshifts. The largest overlap within a single redshift bin is $3\%$ in the highest redshift bin around $z \simeq 0.86$.

\subsection{Halo assignment}
\label{sub:distributing_halos}

As mentioned in section~\ref{sub:intensity_mapping}, the dominant part of HI is expected to be found within galaxies. The challenge of simulating a wide field HI intensity mapping survey is hence that the signal is expected to come from relatively low mass halos that cannot be easily resolved by N-body simulations with the box sizes needed for the sky coverage corresponding to wide field surveys.

Taking a cue from the coarseness of the instrument resolution for HI intensity mapping experiments which erases information about the angular positions of individual sources, one possible prescription for turning density fields into intensity maps is to neglect the discrete origin of the \SI{21}{cm} intensity and to assume that overdensity $\delta$ and brightness temperature fluctuations $\delta T$ are linearly related:
\begin{equation}
	\delta T(z, \mathbf{\Theta}) \approx b_{\rm HI}(z) \bar T(z) \delta(z, \mathbf{\Theta}).
	\label{eq:naiveT}
\end{equation}
In this case, the intensity map would be a rescaled version of the density field. In~\cite{Alonso:2014bq}, for example, the authors followed this approach in their work on creating fast intensity mapping simulations using random log-normal fields. As we expect the HI bias to be closely related to the halo bias, we know that a scale independent, linear bias prescription is only valid at large, linear scales. A possible improvement over equation~\eqref{eq:naiveT} would be a non-linear bias as in~\cite{2013MNRAS.435..743D} which however still ignores possible scale dependencies and the stochastic relation between halo and density field.

We therefore choose to go one step further in exploiting the low resolution of the intensity maps by combining a simulated large scale density field with the conditional mass function of dark matter halos as derived from the halo model. The coarseness of the angular resolution of intensity mapping experiments indeed means that we are not interested in simulating the exact locations of the relevant dark matter halos, but only their total number in each pixel. The conditional mass function $\mathcal N(m|M)$ calibrated by~\cite{2002MNRAS.329...61S}, gives the differential number of halos in the mass range $(m,m+dm)$ within a Lagrangian volume containing mass $M$,
\begin{equation}
		\mathcal N(m|M) dm = \frac M m \frac {|T(s|S)|} {\sqrt{s-S}} \exp\left(- \frac{(B(s) - \delta_{\rm lin}(M))^2} {2 (s - S)}\right) \frac {ds} {s - S}
	\label{eq:cmf}
\end{equation}
where $s$ is the variance of the linear field when smoothed on the Lagrangian scale of the halo mass $m$, $S$ is the corresponding variance at scale $M$, $\delta_{\rm lin}(M)$ is the linearized overdensity corresponding to the mass $M$, and $B(s)$ is the moving barrier shape associated with ellipsoidal collapse:
\begin{equation}
	B(s, z) = \sqrt{a}\delta_{sc}(z)\left(1 + \beta \left(a \frac {\delta_{sc}^2(z)} {s} \right)^{-\alpha} \right),
\end{equation}
where $\delta_{sc}(z)$ is the critical overdensity for collapse at redshift $z$ and the fitting parameters are $a = 0.707$, $\alpha = 0.615$, and $\beta = 0.485$. Finally, $T(s|S)$ is the $5$th order Taylor series of $B(s) - \delta_{\rm lin}(M)$ around $s = S$. 

What we have available from the pixelated density fields is the mass $M$ or equivalently the (non-linear) overdensity $\delta$ contained in each pixel. Ideally we would like to use a mass function conditioned on this $\delta$. In order to apply equation~\eqref{eq:cmf} to the density field, however, we need to relate $\delta$ to its linearized overdensity $\delta_{\rm lin}(M)$. In order to maintain a physical link between $\delta$ and $\delta_{\rm lin}$, we do this by assuming that the density in each pixel approximately follows spherical evolution, in which case we can write~\cite{1994ApJ...427...51B}:
\begin{equation}
	1 + \delta = \left(1 - \frac{\delta_{\rm lin}}{\delta_{sc}}\right)^{-\delta_{sc}}\,,
	\label{eq:deltalin}
\end{equation}
where we have suppressed the mass and redshift dependence. We have tested the approximation by checking that the distribution of $\delta_{\rm lin}$ returned by this procedure is approximately Gaussian for our density fields (skewness $\sim -0.14$ and excess kurtosis $\sim - 0.10$ at the lowest redshift). Going to maps with even smaller redshift bins might be desirable in the future, as radio telescopes can have frequency resolutions that are better than the \SI{25}{\mega\hertz} chosen for our simulations. This would however either require better modeling of $\delta_{\rm lin}$, for example by taking the initial conditions of the simulation into account, or fitting the conditional mass function directly to the non-linear density field.

\begin{figure}[t]
	\centering
	\includegraphics[width = .75\linewidth]{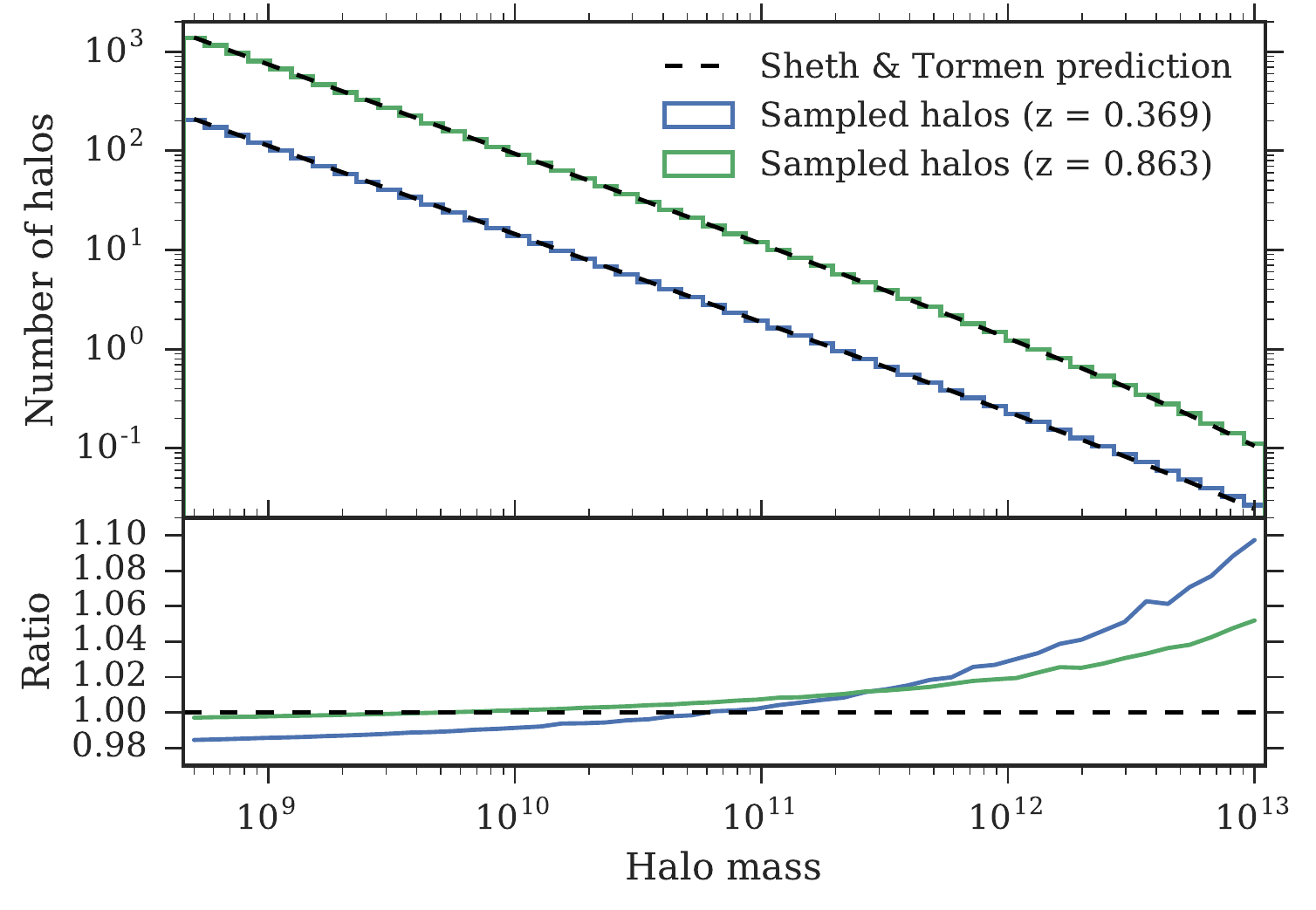}
	\caption{Number of sampled halos within a given mass bin and averaged over all pixels in the lowest (blue) and highest (green) redshift bin. The dashed black lines show the predictions of the unconditional mass function from~\cite{2002MNRAS.329...61S}. The distribution of the sampled halos agrees with the prediction to better than $10\%$.}
	\label{fig:mf_sampled}
\end{figure}

Furthermore, the relation between mass and variance requires us to choose a smoothing filter. The derivation of the mass function from excursion set theory assumes a top-hat filter in Fourier space~\cite{1991ApJ...379..440B}, while halos are typically defined as localized objects in real space. As the results only depend on the variance of the filtered field, standard practice is to use a spherical top-hat filter in real space to relate the halo mass $m$ to the variance $s$. In~\cite{2002MNRAS.329...61S}, also the mass $M$ of the local overdensity is related to a variance $S$ with a spherical top-hat filter. For the pixelated density field, however, a spherical filter is a poor approximation for relating $M$ and $S$, since our pixels have complicated elongated shapes. In order to be consistent, we therefore set $S$ to be the \emph{actual} variance of the linearized pixelated density field across each map. We have checked that the procedure described above leads to an average mass function that is within $10\%$ of the unconditional mass function obtained by setting $S\to0$ and $\delta_{\rm lin}\to0$ in equation~\eqref{eq:cmf} (see Figure~\ref{fig:mf_sampled}).

When populating the density field with halos, we first define a minimum and maximum halo mass. We then Poisson sample the number of halos in each pixel around a mean of 
\begin{equation}
	N(M) = \int_{m_{\rm min}}^{m_{\rm max}}dm\, \mathcal N(m|M),
	\label{eq:nofm}
\end{equation}
where $M$ is the mass in the pixel. Finally, we sample a mass $m$ for each of the halos from the distribution\footnote{In principle, one might worry that this procedure does not conserve mass~\cite{1999MNRAS.304..767S}; in our case, however, we have $m_{\rm max}\ll M$ and we have checked that mass conservation is not a problem in practice.} $\mathcal N(m|M)/N(M)$. We end up with a sample of halo masses that is drawn from the conditional mass function for each individual pixel on the map. 

When following the described procedure, we however ignore the effect of peculiar velocities on the simulated maps. Peculiar velocities of the halos can shift the \SI{21}{cm} line of the HI within some of the halos into neighboring frequency bins, thereby distorting the line-of-sight boundaries between the pixels. As the prospects of redshift space distortion (RSD) measurements with the \SI{21}{cm} line are promising~\cite{2015aska.confE..31R,2015ApJ...803...21B}, we propose a way for including them in the simulations in Appendix~\ref{sub:peculiar_velocities}.

\subsection{HI intensity maps}
\label{sub:intensity_maps}

To simulate the sky as seen by a HI intensity mapping survey, we need to turn the distribution of halos and their masses into a distribution of HI mass. For a summary of different approaches to this problem see~\cite{Padmanabhan:2014ug} and~\cite{Padmanabhan:2015vg}. Our approach closely follows work in~\cite{Bagla:2010fu}: For halos that satisfy a constraint on the circular velocity of the halo, we assign a fraction of the halo mass as HI mass. As proposed in~\cite{Padmanabhan:2015vg}, we use a halo mass to HI mass ratio $\alpha = 0.15$ together with an exponential cut-off at circular velocities $v_{c,0} = \SI{30}{km.\per.s}$ and $v_{c,1} = \SI{200}{km.\per.s}$:
\begin{equation}
	m_{\rm HI}(m) = \alpha (1 - Y_p) \frac {\Omega_b} {\Omega_m}  m \exp\left(-\left(\frac{v_{c,0}} {v_c(m)}\right)^3\right) \exp\left( - \left(\frac{v_c(m)} {v_{c,1}}\right)^3\right),
	\label{eq:mh1m}
\end{equation}
where $m$ is again the mass of the halo. $\Omega_b$ and $\Omega_m$ are set by the simulations and we use a Helium fraction of $Y_p = 0.24$. Using equation~\eqref{eq:mh1m} to assign a HI mass to all the halos in a given pixel, the total HI mass in the pixel is simply given by the sum over all halos. We can finally rescale the resulting HI mass map into an intensity map of \SI{21}{cm} brightness temperature by using equation~\eqref{eq:tbar}.

\begin{algorithm}[t]
	\begin{algorithmic}[1]
	\State{create mass map of matter distribution from N-body lightcone output}
	\For{mass $M$ in pixel $i$ of the mass map}
	\State turn $M$ into an overdensity $\delta$
	\State turn $\delta$ into a linearized overdensity $\delta_{lin}$ using eq.~\eqref{eq:deltalin}
	\State calculate $\mathcal N(m|M)dm$ using eq.~\eqref{eq:cmf}
	\State sample the number of halos $n$ from a Poisson distribution centered on $N(M)$, eq.~\eqref{eq:nofm}
	\State sample $n$ halo masses $\{m_j\}_{j=1,\cdots,n}$ from $\mathcal N(m|M)/N(M)$ using inversion sampling
	\State turn $\{m_j\}$ into HI masses $\{m_{{\rm HI},j}\}$ using eq.~\eqref{eq:mh1m}
	\State set pixel $i$ of HI mass map to $\sum_jm_{{\rm HI},j}$
	\EndFor
	\State{turn HI mass map into \SI{21}{cm} intensity map using eq.~\eqref{eq:tbar}}
	\end{algorithmic}
	\caption{Algorithm for simulating HI intensity maps.}
	\label{alg:alg}
\end{algorithm}

\begin{figure}[t]
	\centering
	\includegraphics[width = .75\linewidth]{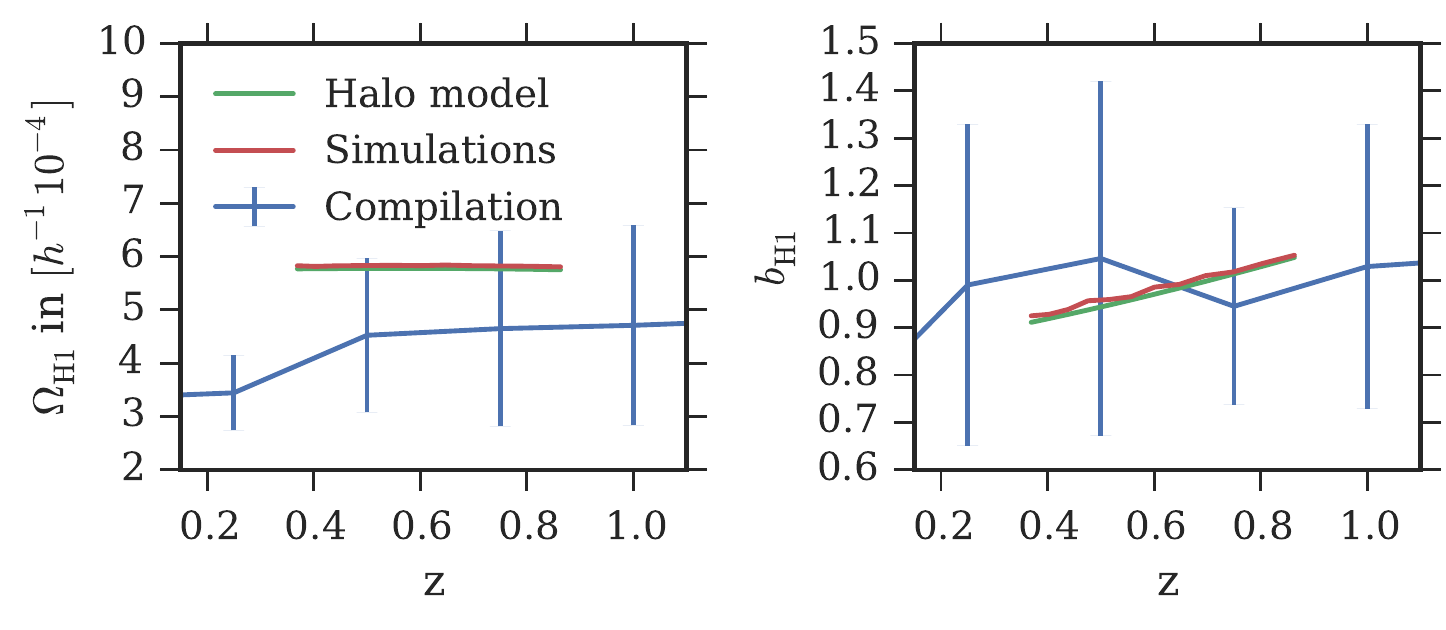}
	\caption{Comparison of density (left panel) and large scale biasing of HI (right panel) as a function of redshift as predicted by the unconditional mass function from~\cite{2002MNRAS.329...61S} (green) for the prescription of eq.~\eqref{eq:mh1m}, as measured on our simulated intensity maps (red), and from the compilation in~\cite{Padmanabhan:2014ug} (blue). The values from the unconditional halo model agree well with the simulations by construction, with small deviations coming from the imperfect linearization (equation \eqref{eq:deltalin}). The $\Omega_{\rm HI}$ values from our prescription are slightly high with respect to the compilation in~\cite{Padmanabhan:2014ug}. Once better data is available, a more significant mismatch in amplitude could be easily fixed by adjusting $\alpha$ in equation~\eqref{eq:mh1m}.}
	\label{fig:omegah1andbias}
\end{figure}

Algorithm~\ref{alg:alg} gives a summary of the proposed procedure for generating HI intensity maps. For the dark matter halos, the main assumptions of our procedure are that the linearized field is well approximated by equation~\eqref{eq:deltalin} (line 4) and that the conditional mass function sampling within our pixels (line 5 and 6) is a good description of the halo distribution. In Appendix~\ref{sec:sub_grid} we study those steps in more detail. Our main arguments for the validity of our approach are the good agreement of the resulting unconditional halo mass function (see Figure~\ref{fig:mf_sampled}) and large scale bias (see Figures~\ref{fig:omegah1andbias} and~\ref{fig:mattervssimcells}) with the analytical results from~\cite{2002MNRAS.329...61S}. The analytic expressions have been shown to be in reasonable agreement with N-body simulations~\cite{2007MNRAS.374....2R,2010MNRAS.403.1353C,2013MNRAS.433.1230W,2015MNRAS.450.1349K,2015arXiv150506436H}, and the fact that we are reproducing the analytical results at the mass function and bias level is encouraging. We leave more detailed investigations of the accuracy of our procedure with high-resolution simulations of smaller volumes for future work.

\begin{figure}[t]
	\centering
	\includegraphics[width = .75\linewidth]{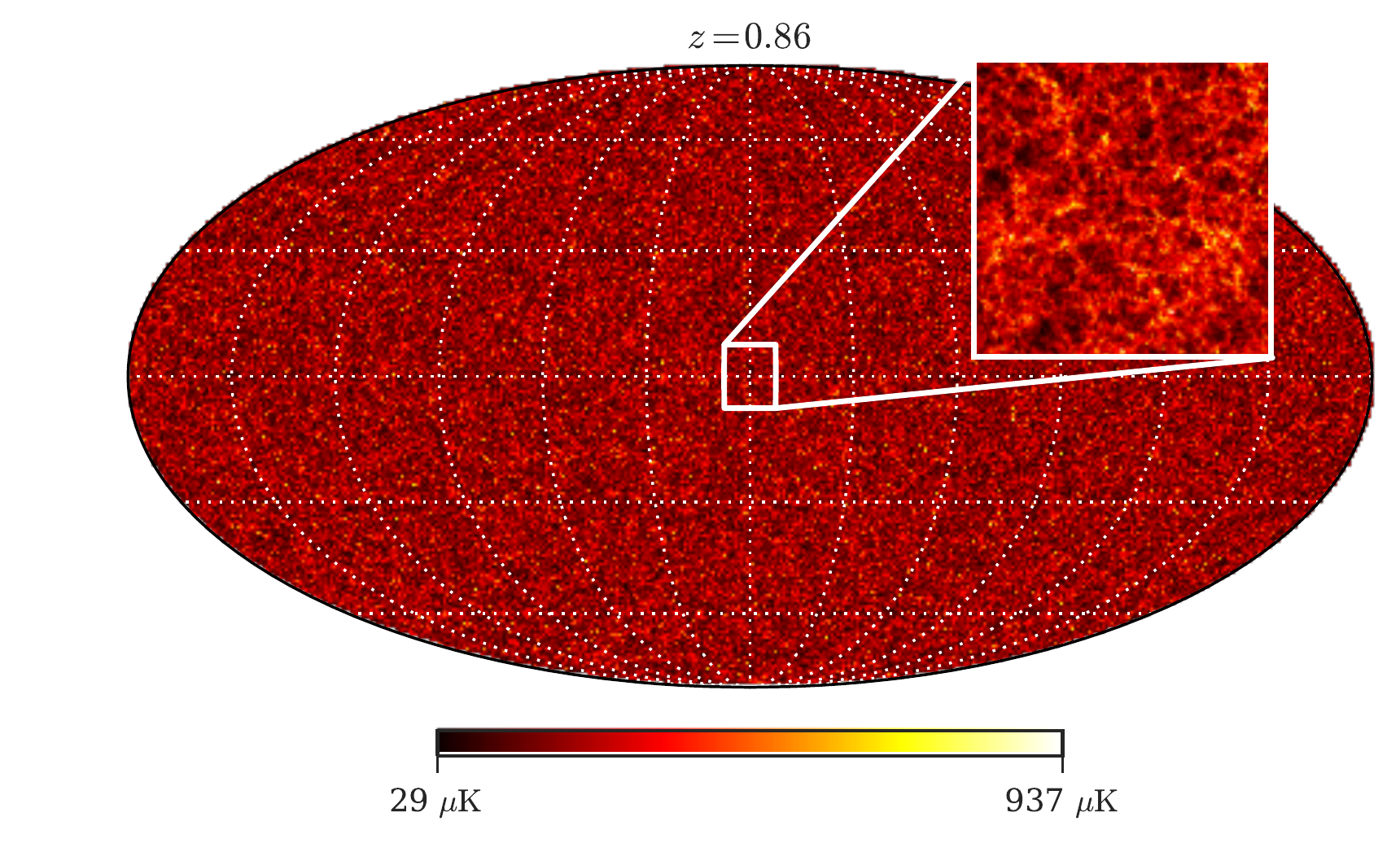}
	\includegraphics[width = .75\linewidth]{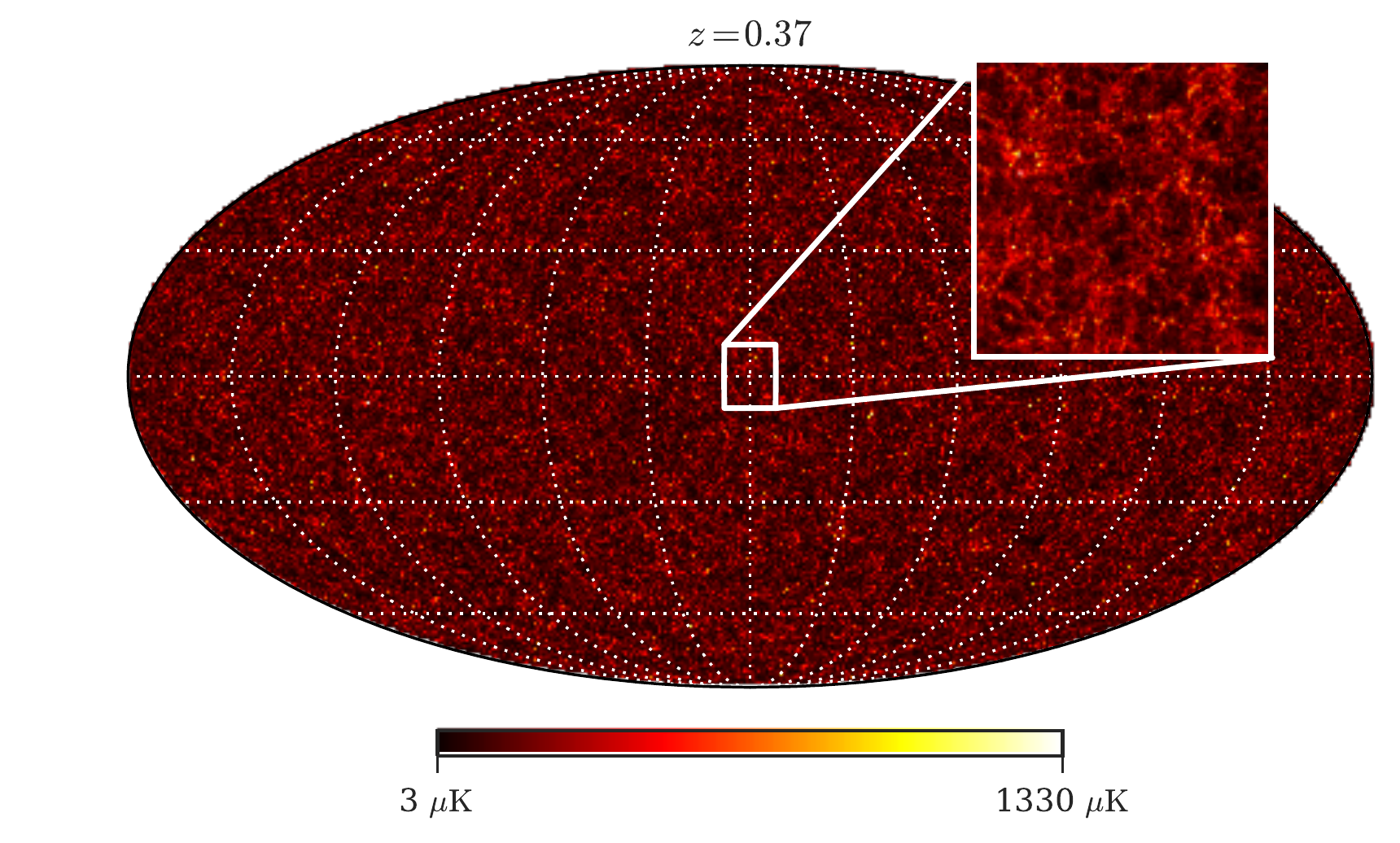}
	\caption{Simulated HI intensity maps in units of \si{\micro\kelvin} brightness temperature at $z \simeq 0.863$ (top) and $z \simeq 0.369$ (bottom). The zoom regions have side lengths of \ang{15}.}
	\label{fig:intensitymap}
\end{figure}

\begin{figure}[t]
	\centering
	\includegraphics[width = .75\linewidth]{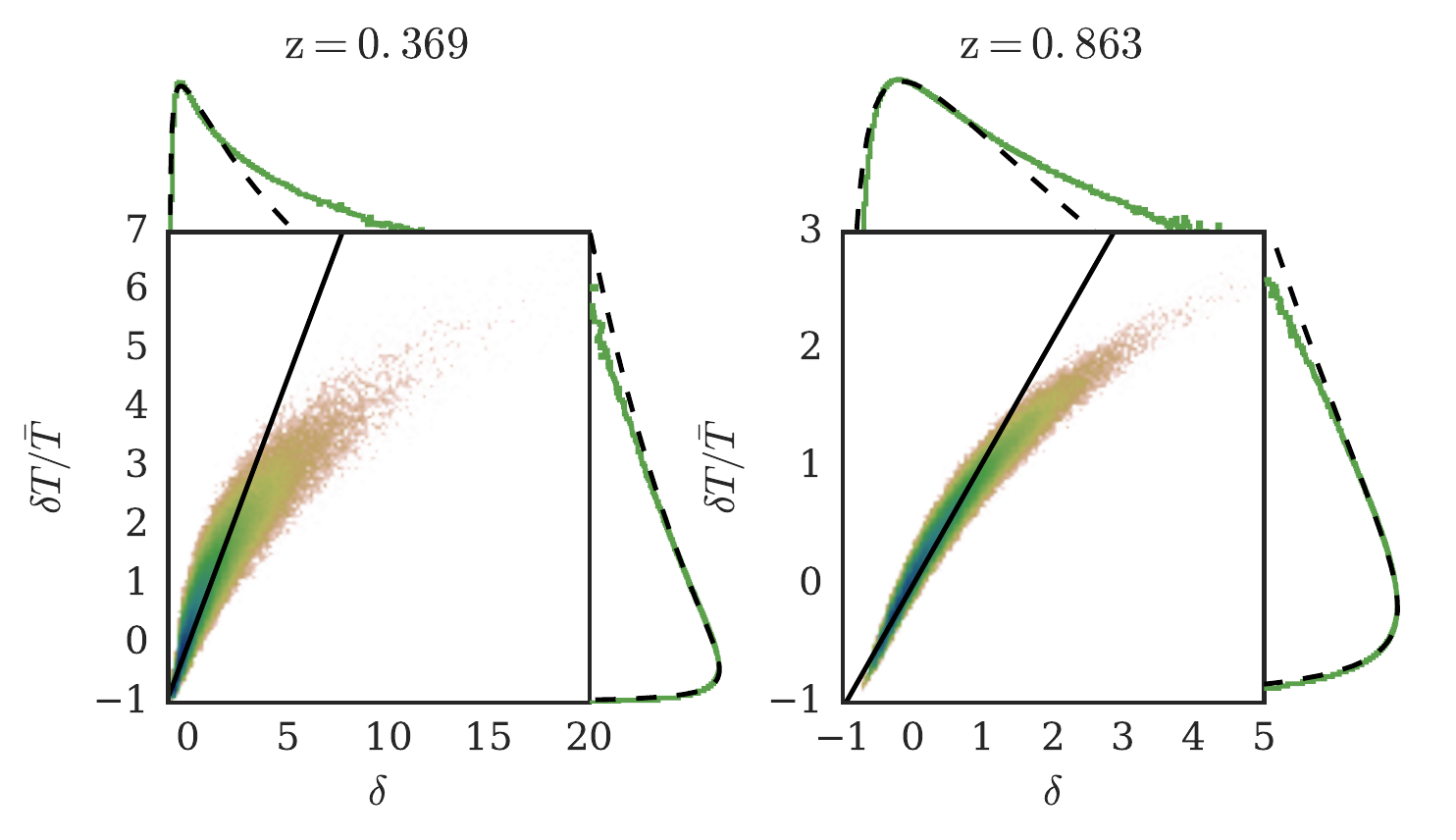}
	\caption{Relation between matter overdensity $\delta$ and temperature fluctuations $\delta T/\bar T$ in the simulated maps at redshift $z \simeq 0.863$ (right) and $z \simeq 0.369$ (left). The one-dimensional histograms show the individual distributions of $\delta$ and $\delta T/\bar T$ along with a log-normal distribution with the same mean and variance (black line). The two-dimensional histograms show the non-linear, stochastic relation between $\delta$ and $\delta T/\bar T$ as compared to a simple linear bias model (black line). Both the color-scaling of the two-dimensional histograms and the y-axis of the one-dimensional histograms are log-scaled.}
	\label{fig:deltavsdt}
\end{figure}

Figure~\ref{fig:intensitymap} shows the simulated HI intensity maps at the lowest and highest redshift ($z \simeq 0.369, 0.863$, respectively). In Figure~\ref{fig:deltavsdt}, we show the relation between dark matter overdensity $\delta$ and brightness temperature fluctuations $\delta T$ induced by the conditional halo mass function sampling. Similar to findings in studies with high resolution N-body simulations (see e.g.~\cite{2011MNRAS.415..383M}), we find that the relation between halo overdensity and matter overdensity deviates from a linear behavior for large overdensities. Figure~\ref{fig:deltavsdt} also shows the individual distributions of $\delta$ and $\delta T$ together with a log-normal distribution of the same mean and variance.

The left panel of Figure~\ref{fig:omegah1andbias} shows a comparison of $\Omega_{\rm HI}(z)$ resulting from our prescriptions along with the data points of the compilation in~\cite{Padmanabhan:2014ug} consisting of damped Lyman-$\alpha$ and intensity mapping observations. We can see that equation~\eqref{eq:mh1m} yields a somewhat high $\Omega_{\rm HI}$ compared to the data compilation, as the prescription was fitted to another data compilation and for a cosmology with a lower $\sigma_8$. $\Omega_{\rm HI}$ could be matched more precisely by adjusting the parameter $\alpha$ in equation~\eqref{eq:mh1m} which would simply result in an overall rescaling of all maps. As all our results are independent of the overall amplitude, we decided to simply adopt the value of $\alpha$ proposed in~\cite{Padmanabhan:2015vg}. The mapping between halo mass and HI mass will not be one-to-one in reality, and more work has to go into the question on how much coarse graining is allowed when relating halo and HI mass as more data becomes available. A more detailed modeling of e.g. the scatter around this relation is however beyond the scope of this work.

\section{Angular power spectrum}
\label{sec:the_angular_power_spectrum}

In this section, we present our results on the angular clustering of the simulated HI intensity maps as described by their angular power spectrum. We first describe the estimators that we apply to our intensity maps and then discuss how the non-linearities affect the angular power spectrum. 
We want to study how the estimators are affected by the non-linear, non-Gaussian field. The non-linearities are strongest at low redshifts, and we will in the following focus on the auto-correlation spectrum $C_\ell(z)$ in the lowest redshift bin. We contrast the results with the auto-correlation spectrum in the highest redshift bin to compare the performance for different levels of non-Gaussianity.

\subsection{Estimation}
\label{sub:estimation_of_angular_power_spectra}

Given a map with brightness temperatures $T_i$ drawn from a field with underlying angular power spectrum $C_\ell$ and its coefficients $a_{lm}$ of the map's expansion in spherical harmonics, the angular power spectrum can be estimated as
\begin{equation}
	\hat C_\ell = \frac 1 {2\ell + 1} \sum_{m = -\ell}^{\ell} \left|a_{lm}\right|^2.
	\label{eq:naivecell}
\end{equation}
When only parts of the sky are available, recovering the spherical harmonics from the masked field is more complicated. The pseudo $C_{\ell}$ estimator~\cite{1973ApJ...185..413P} is a standard approach to this problem: In this approximation, the masked pixels are set to the mean value of the unmasked part before the full sphere is decomposed in spherical harmonics. The power spectrum is then estimated via
\begin{equation}
	\hat C^{\rm pseudo}_\ell = \frac 1 {(2\ell + 1)f_{\rm sky}} \sum_{m = -\ell}^{\ell} \left|a_{\ell m}\right|^2
\end{equation}
with $f_{\rm sky}$ being the fraction of unmasked sky~\citep[see also][for more sophisticated approaches]{1997PhRvD..55.5895T, 1998PhRvD..57.2117B, Szapudi:2000vt, 2002ApJ...567....2H, Efstathiou:2004bh}. 

In addition to the pseudo $C_\ell$ estimator, we also consider the publicly available package PolSpice~\cite{2001ApJ...561L..11S, Chon:2004ki} in our analyses. PolSpice first estimates the correlation function of the masked fluctuations, corrects for the mask, and then calculates the demasked angular power spectrum from the corrected correlation function. To avoid artifacts from the decomposition of an incomplete correlation function of the masked field, the correlation function has to be smoothed with an apodization function. We used Gaussian random fields with a mask corresponding to those applied to the simulated intensity maps to fix the parameters of a Gaussian apodization (width of $\sigma = 90^\circ$ and a maximum angle of the correlation function $\beta_{max} = 120^\circ$).

\subsection{Results}
\label{sub:results_spec}

\begin{figure}[t]
	\centering
	\includegraphics[width = .75\linewidth]{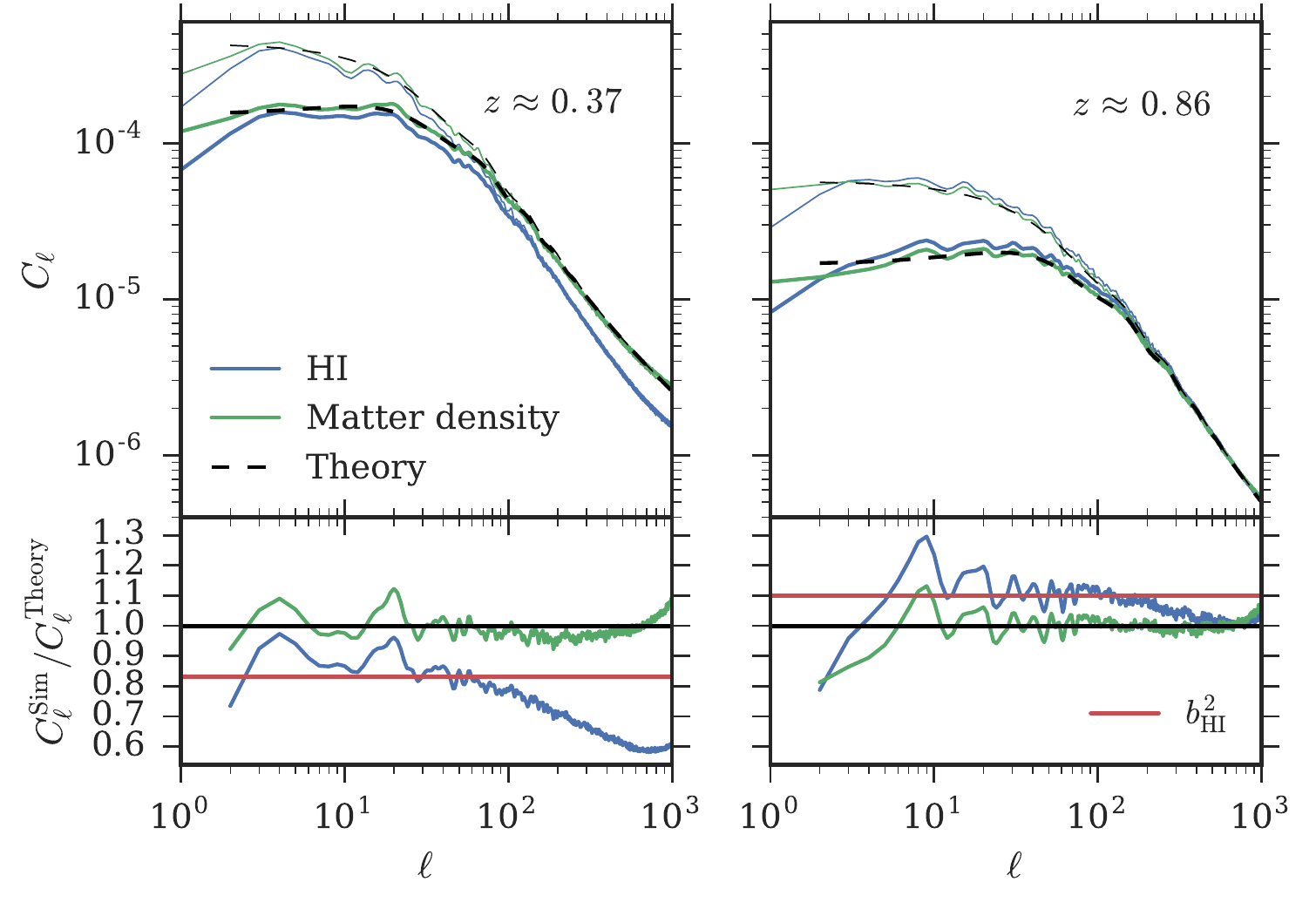}
	\caption{Angular power spectrum as estimated from the matter density (green) and HI intensity maps (blue) using PolSpice, averaged over all realizations of the simulations. We show the dimensionless spectrum for the intensity maps, scaled by the mean brightness temperature. The theory prediction (black) for the matter power spectrum are calculated with CLASS~\cite{Blas:2011jb, Dio:2013ia} and halofit~\cite{Smith:2003ui, Takahashi:2012ir}. The thick lines show the spectra when ignoring redshift space distortions, while the thin lines are showing the results for the distorted maps (see Appendix~\ref{sub:peculiar_velocities} for more details on our modeling of redshift space distortions). The bottom panel shows the ratio between the predicted matter spectrum and the estimated spectra for the matter and intensity maps without redshift space distortions. As expected, the spectrum of the matter field is consistent with the prediction and their ratio is close to one. Deviations after $\ell \simeq 700$ are due to pixel effects. The ratio of intensity mapping spectra and theory prediction for the matter field should be compared to the large scale HI bias from equation~\eqref{eq:bHI-halomodel} (red). We see that the HI bias matches the ratio at large scales ($\ell \lesssim 50$) and that there is a scale dependence at smaller scales.}
	\label{fig:mattervssimcells}
\end{figure}

In Figure~\ref{fig:mattervssimcells}, we show the angular power spectrum as estimated from the matter density and HI intensity maps using PolSpice, averaged over all 40 realizations. We also show the theory predictions for the angular power spectrum of the matter density field from equation~\eqref{eq:lintheory} as calculated with CLASS~\cite{Blas:2011jb, Dio:2013ia}, using halofit~\cite{Smith:2003ui, Takahashi:2012ir} to model the non-linearities. The bottom panel shows the ratio of the estimated to the predicted power spectra. Figure~\ref{fig:mattervssimcells} also shows the angular power spectra of the maps when including redshift space distortions (thin lines, see Appendix~\ref{sub:peculiar_velocities} for details on our modeling of peculiar velocities). In linear theory, the redshift space distortions lead to a modification of the power spectrum~\cite{1987MNRAS.227....1K} that results in an enhancement of the angular power spectrum at large scales but does not affect scales much smaller than the size of the redshift bin (see e.g.~\cite{2007MNRAS.378..852P,2011PhRvD..84f3505B} for more details). Figure~\ref{fig:mattervssimcells} shows that the linear theory behavior of the redshift space distortions as calculated with CLASS is correctly reproduced by our distorted maps.

As expected, the matter power spectrum is consistent with the theory prediction over most scales with a deviation at small scales ($\ell > 700$) due to pixel effects. The effect of non-linearities in the density field at the power spectrum level is hence well modeled by the halofit corrections to the linear power spectrum. The agreement also confirms that, to the precision we are interested in, the overlap of a few percent per mass map within the simulated box (see also section~\ref{sub:simulating_density_fields}) is not affecting the angular power spectra.

The ratio between the intensity mapping spectra and the predicted matter density is seen to be close to $b_{\rm HI}^2$ on large scales ($\ell \lesssim 50$). Even though the agreement with the analytic prediction for the large scale bias is expected, it is a non-trivial consequence of the conditioning of the mass function on the overdensities within each pixel. Figure~\ref{fig:mattervssimcells} hence shows that our prescription for assigning halos to the density field reproduces the halo statistics at the two-point level on large scales remarkably well. On smaller scales, our modeling of the halo distribution with the conditional mass function yields a mildly scale dependent bias in particular at the lowest redshift, where the ratio falls by $\sim 10\%$ of its large scale value at $\ell \approx 200$. The right panel of Figure~\ref{fig:omegah1andbias} shows a comparison of $b_{\rm HI}$ from equation~\eqref{eq:bHI-halomodel} with the large-scale bias in the simulated intensity maps, estimated as the square root of the average ratio of the intensity mapping angular power spectra and the theory predictions on scales $10<\ell<50$. Figure~\ref{fig:omegah1andbias} also shows good agreement with a compilation of bias values in~\cite{Padmanabhan:2014ug}, derived from a selection of theoretical prescriptions from the literature.

\section{Covariance of the angular power spectrum}
\label{sec:covariance_of_angular_power_spectra}

For precise cosmological inference from the angular power spectrum, it is important to know the distribution of its estimator. To first order, this means that we need to estimate the covariance of the angular power spectrum as measured from intensity maps. As in section~\ref{sec:the_angular_power_spectrum}, we first describe our estimation procedure before discussing the results, focusing in particular on the effect of non-linearities on the covariance of the angular power spectrum.

\subsection{Estimation}
\label{sub:cov_estimation}

We describe and compare three standard approaches to the problem of covariance estimation: estimation from multiple simulations, estimators based on Gaussian random fields and a jackknife approach.

\subsubsection{Multiple simulations}

As already mentioned in section~\ref{sub:simulating_density_fields}, we use the density field from 10 independent N-body simulations as the basis for our intensity maps. We use each of the boxes to create 4 simulations which cover a quarter of the sky. If we focus on the first quadrant of each simulation, then we estimate the covariance in this case as:
\begin{equation}
	{\rm Cov}(C_{\ell}, C_{\ell'}) = \frac {1} {n - 1} \sum_{i = 0}^{n} (C_{\ell}^{i} - \bar C_{\ell}) (C_{\ell'}^{i} - \bar C_{\ell'}),
\label{eq:covrepeat}
\end{equation}
where $n = 10$, $C_{\ell}^{i}$ is the power spectrum estimated from the $i^{\rm th}$ realization (this could be either the pseudo $C_\ell$ or PolSpice estimator), $\bar C_{\ell} = (1/n)\sum_{i = 0}^{n} C_{\ell}^{i}$ is the mean over all realizations and the $n - 1$ denominator is chosen to yield an unbiased estimator of the underlying covariance. We do this for each quadrant and take the final covariance matrix estimator to be the arithmetic mean of ${\rm Cov}(C_{\ell}, C_{\ell'})$ over all four quadrants. This way we avoid that correlations between the four quadrants introduced through the finite size of the simulated box affect the estimated covariance. In Appendix \ref{sub:peculiar_velocities}, we additionally study the effect of redshift space distortions on the estimated covariances.

\subsubsection{Gaussian approximation}

In the case of an isotropic Gaussian random field on a full sphere, the coefficients $a_{lm}$ of the spherical harmonic decomposition are independent Gaussian random variables and identically distributed for each $\ell$. Consequently, the estimator $\hat C_{\ell}$~\eqref{eq:naivecell} is chi-squared distributed with mean $C_\ell$ and variance \citep{1995PhRvD..52.4307K}
\begin{equation}
	{\rm Var}(\hat C_\ell) = \frac 2 {2 \ell + 1} C_\ell^2.
\end{equation}
The covariance ${\rm Cov}(C_\ell,C_{\ell'})$ is zero for $\ell \neq \ell'$ in this case.

As for the estimation of the $C_\ell$s, these considerations get more involved in the case of partial sky coverage (see e.g.~\cite{1997PhRvD..55.5895T, 1998PhRvD..57.2117B, Szapudi:2000vt, 2002ApJ...567....2H, Efstathiou:2004bh, Cabre:2007jq}). A standard approximation for the variance of the pseudo $C_{\ell}$ estimator is given by~\cite{2002ApJ...567....2H}
\begin{equation}
	{\rm Var}(\hat C^{\rm pseudo}_\ell) \approx \frac 2 {(2 \ell + 1)f_{\rm sky}} C_\ell^2,
	\label{eq:pseudovar}
\end{equation}
where as before $f_{\rm sky}$ is the fraction of unmasked sky and ${\rm Cov}(\hat C^{\rm pseudo}_\ell,\hat C^{\rm pseudo}_{\ell'})$ is approximated to be zero for $\ell \neq \ell'$. Since this approximation neglects correlations between individual $\ell$ scales as introduced by the mask, it is a good approximation only for bandpowers
\begin{equation}
	B_\ell = \frac 1 {2\Delta\ell + 1} \sum_{\ell' = \ell-\Delta\ell}^{\ell+\Delta\ell} C_{\ell'},
	\label{eq:bandpowers}
\end{equation}
where $\Delta\ell$ is chosen such that the correlations between the bandpowers is small. Following the definition of $B_\ell$, the variance of the bandpowers relates to the $C_\ell$ variance as follows:
\begin{equation}
	{\rm Var}(B_\ell) = \frac 1 {(2\Delta\ell + 1)^2} \sum_{\ell' = \ell-\Delta\ell}^{\ell+\Delta\ell} \sum_{\ell'' = \ell-\Delta\ell}^{\ell+\Delta\ell} {\rm Cov}(C_{\ell'},C_{\ell''}).
	\label{eq:bperror}
\end{equation}

Besides the pseudo $C_{\ell}$ approach, we will also use the PolSpice estimator for the covariance~\citep{Efstathiou:2004bh} which is based on Gaussian field statistics and takes the mask into account. We calculate the Gaussian field estimators for each simulation individually and report the results in section~\ref{sub:results_cov}.

\subsubsection{Jackknife estimation}

Resampling techniques aim at estimating the covariance of an estimator by manipulating the sample on which it is based. The advantage of resampling techniques is that they do not assume a field with specific statistical properties. The disadvantage is that they typically assume independence of different parts of the sample which is not satisfied by correlated fields. In a jackknife approach, one studies the behavior of an estimator when parts of the data are ignored in the estimation process. In the context of our maps this implies that we study the distribution of angular power spectra as derived from maps where a subset of the simulated brightness temperature pixels is masked. We split the simulated part of the sky into $n_{\rm jack}$ parts and estimate all power spectra $C_{\ell}^{{\rm jack}, i}$ (using either the pseudo $C_\ell$ or PolSpice estimators) when the $i^{\rm th}$ of the $n_{\rm jack}$ parts is masked out. The covariance of the angular power spectrum is then estimated as
\begin{equation}
	\text{Cov}_{\rm jack}(C_{\ell}, C_{\ell'}) = \frac {n_{\rm jack} - 1} {n_{\rm jack}} \sum_{i = 1}^{n_{\rm jack}} (C_{\ell}^{{\rm jack}, i} - \bar C_{\ell}^{\rm jack}) (C_{\ell'}^{{\rm jack}, i} - \bar C_{\ell'}^{\rm jack})
\label{eq:covjack}
\end{equation}
with $\bar C_{\ell}^{\rm jack} = (1/n_{\rm jack}) \sum_{i = 1}^{n_{\rm jack}} C_{\ell}^{{\rm jack}, i}$. In our analysis, we define the masked-out patches as pixels in coarser HEALPix maps. The size of the patches was determined after an analysis of stochastic fields and is discussed in Appendix~\ref{sec:toy_models}. As for the Gaussian field estimate, we calculate a jackknife estimate individually for each simulation and report the results in section~\ref{sub:results_cov}.

\subsection{Results}
\label{sub:results_cov}

To present our results on the covariance of angular power spectra of HI intensity maps, we decompose the covariance into its diagonal part (the variance) and its off-diagonal part (the correlation matrix) as given by:
\begin{align}
	\sigma^2(C_\ell) &= \text{Cov}(C_{\ell}, C_{\ell}),\\
	\text{Corr}(C_{\ell}, C_{\ell'}) &= \frac {\text{Cov}(C_{\ell}, C_{\ell'})} {\sigma(C_\ell)\sigma(C_{\ell'})}.
\end{align}
For simplicity, we will call the results regarding the covariance estimated from the 10 independent realizations of the intensity maps the \emph{simulation estimator} in the following. We present a covariance analysis for Gaussian and log-normal random fields with similar statistical properties as the simulated intensity maps and for our survey geometry of a contiguous quarter of the sky in Appendix~\ref{sec:toy_models}.

\begin{figure}[t]
	\centering
	\includegraphics[width = .74\linewidth]{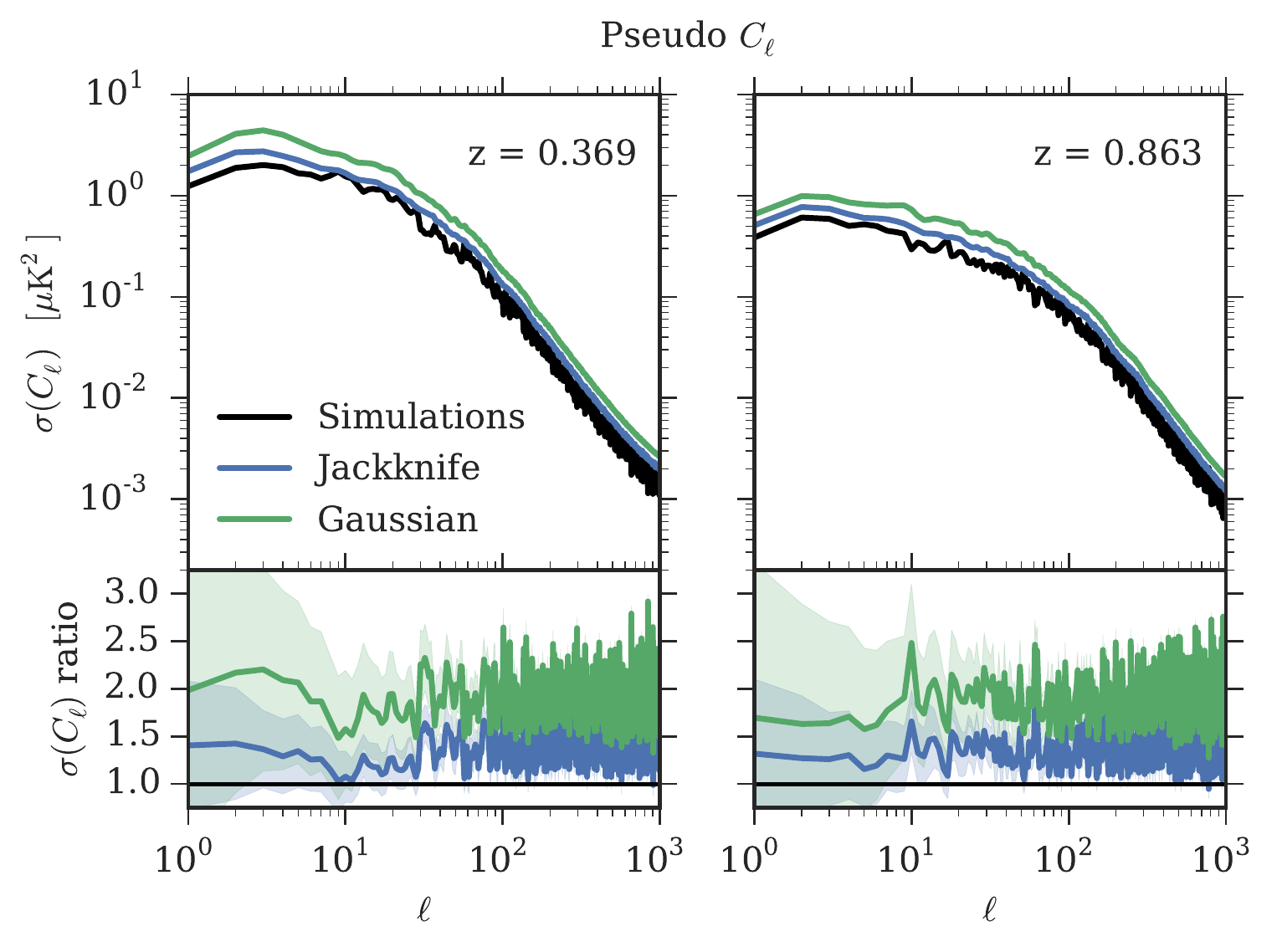}
	\includegraphics[width = .74\linewidth]{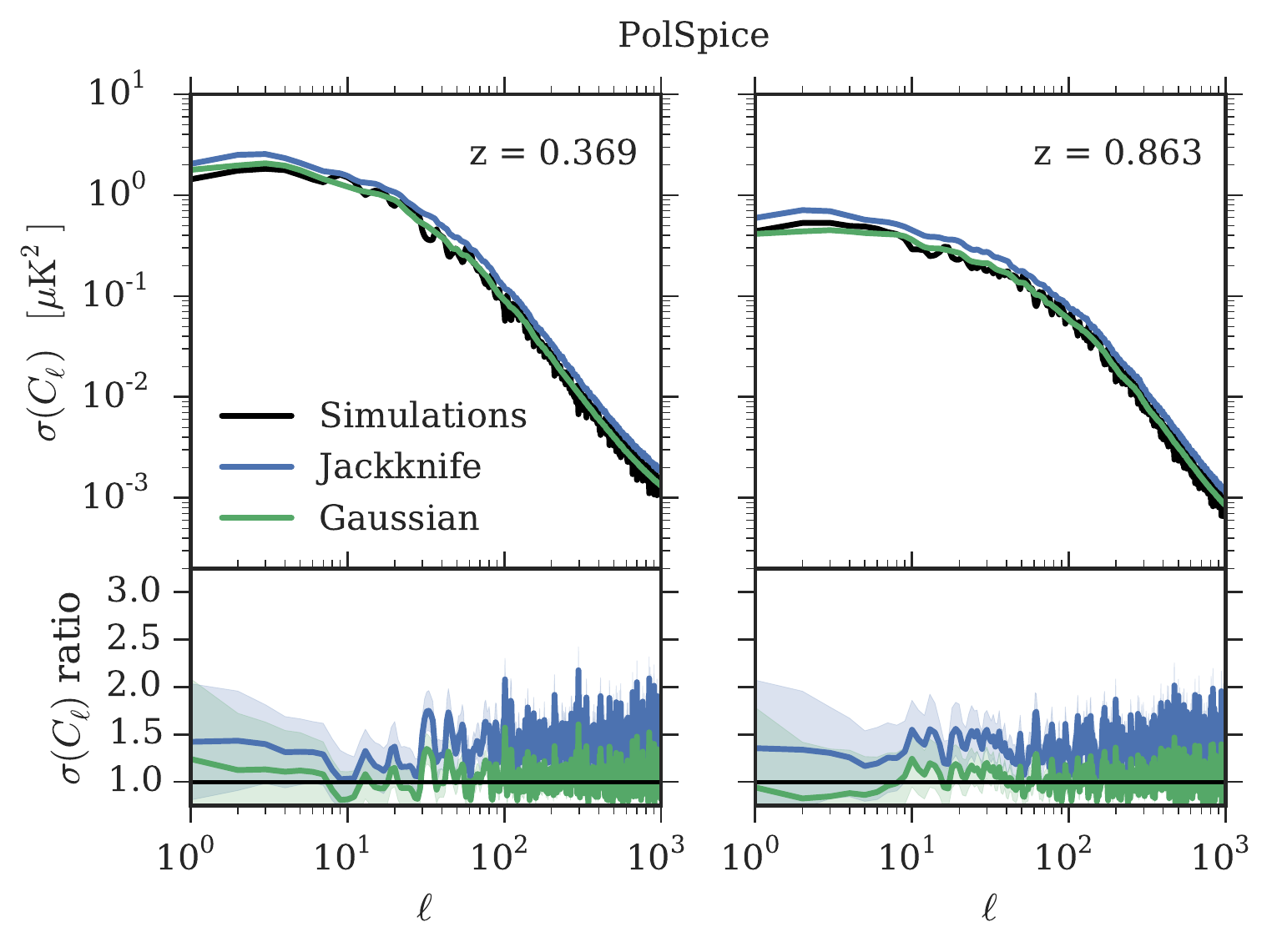}
	\caption{Standard deviation $\sigma(C_\ell)$ of pseudo $C_\ell$ (top) and PolSpice (bottom) estimators as estimated from the 10 independent simulations (black), the jackknife (blue), and the Gaussian assumption (green) for the lowest ($z \simeq 0.369$, left panel) and highest ($z \simeq 0.863$, right panel) redshift bin of the simulations. The bottom panel of each plot shows the ratio of the jackknife and Gaussian estimator to the result from the 10 realizations. The shaded areas in the bottom panels show the standard deviation of the ratio, estimated from the distribution of the jackknife and Gaussian estimator over the 10 realizations. For the pseudo $C_\ell$ estimator, $\sigma(C_\ell)$ is overpredicted by the Gaussian estimator by a factor of $\sim 2$. Within the noise, the PolSpice estimate is consistent with the result from the independent simulations. The jackknife estimate overestimates $\sigma(C_\ell)$ by about $40\%$, independently of scale, redshift, and estimator.}
	\label{fig:h1_cellerr}
\end{figure}

\subsubsection{Variance}

\begin{figure}[t]
	\centering
	\includegraphics[width = .75\linewidth]{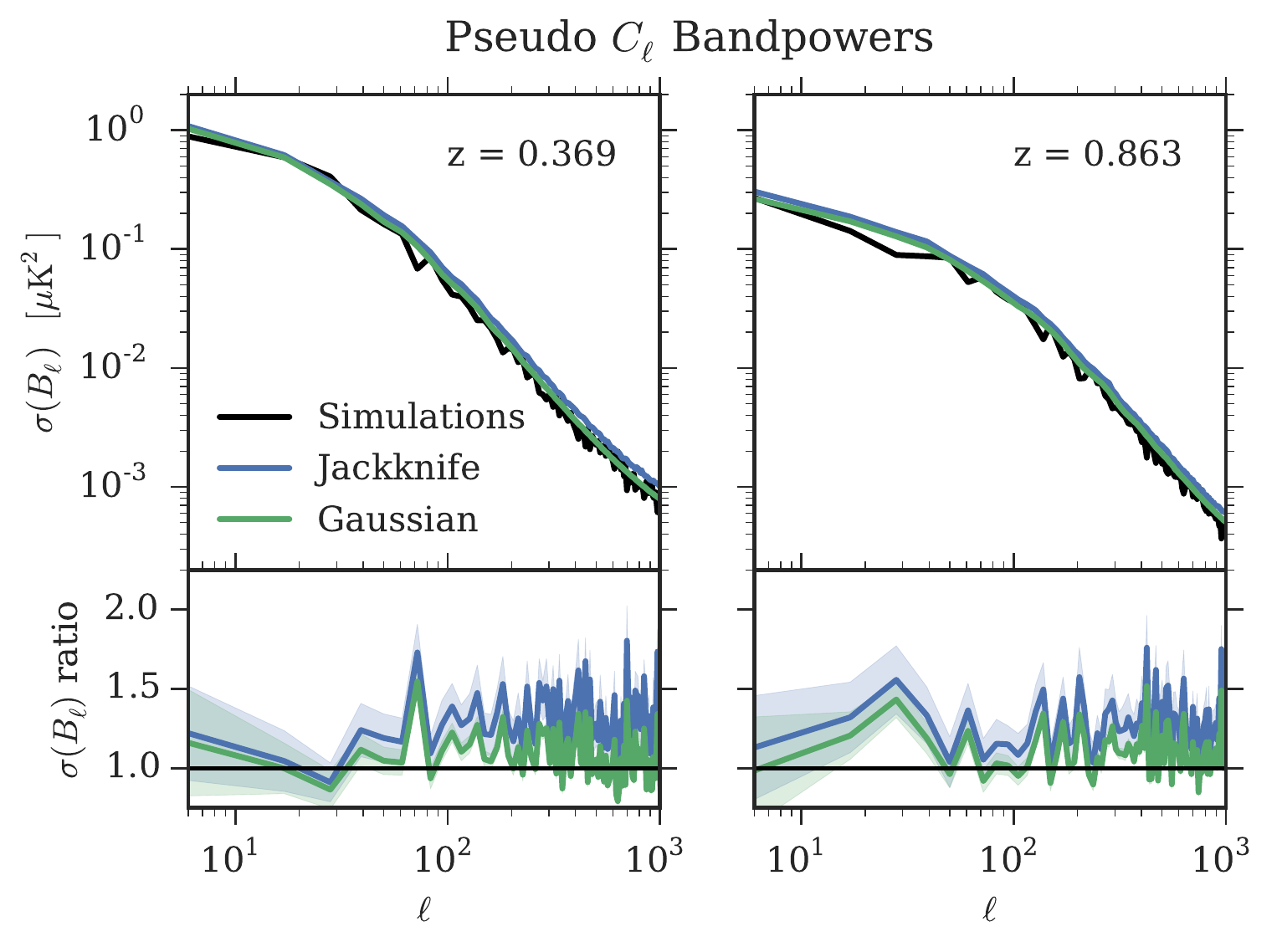}
	\caption{Standard deviation $\sigma(B_\ell)$ of pseudo $C_\ell$ bandpowers defined in equation \eqref{eq:bandpowers} as estimated from the 10 independent simulations, the jackknife, and the Gaussian assumption for the lowest ($z \simeq 0.369$, left panel) and highest ($z \simeq 0.863$, right panel) redshift bin of our simulations. The bottom panel shows the ratio of the jackknife and Gaussian estimator to the result from the 10 realizations. The shaded areas in the bottom panels show the standard deviation of the ratio, estimated from the distribution of the jackknife and Gaussian estimator over the 10 realizations. The bandpowers are averaged over a bandwidth of $\Delta\ell = 5$. The bias of the pseudo $C_\ell$ estimate for $\sigma(B_\ell)$ is dropping from $90\%$ in the case of the unbinned spectrum (Figure \ref{fig:h1_cellerr}) to below $10\%$ for the bandpowers.}
	\label{fig:h1_bperr}
\end{figure}

Figure~\ref{fig:h1_cellerr} shows the different estimates of $\sigma(C_\ell)$ for the lowest ($z \simeq 0.369$) and highest ($z \simeq 0.863$) redshift bin of the intensity map simulations presented in section \ref{sec:simulating_h_i_intensity_maps}. For all redshifts and estimators, the jackknife estimator~\eqref{eq:covjack} is consistently biased\footnote{Whenever we talk about bias in this section, we refer to the bias of an estimator in the statistical sense, i.e. as a mismatch between expected value of the estimator and the true, underlying quantity. This is not to be confused with the concept of halo and HI bias discussed in sections \ref{sub:large_scale_structure_of_hi} and \ref{sub:results_spec}.} to higher $\sigma(C_\ell)$ by approximately $\sim 40\%$, an effect that is also seen for the stochastic field models in Appendix~\ref{sec:toy_models}. The PolSpice estimator is consistent with the simulation estimator within the statistical noise. We find that the pseudo $C_\ell$ estimate for $\sigma(C_\ell)$ given in equation~\eqref{eq:pseudovar} predicts a diagonal error on the angular power spectrum that is a factor of $\sim 2$ higher than the simulation estimator, compensating for the lack of correlations (see Figure~\ref{fig:h1_cellerr}). To account for this effect, we also show the results for the variance of pseudo $C_\ell$ bandpowers $B_\ell$ for a bandwidth of $\Delta \ell = \pm 5$ (see equation~\ref{eq:bandpowers}). As shown in Figure~\ref{fig:h1_bperr}, the bias of the pseudo $C_\ell$ estimate for $\sigma(B_\ell)$ drops below $10\%$. We have checked that this bias decreases further when increasing $\Delta\ell$ at the cost of losing more and more information due to the averaging over $\ell$. 

Overall, the variance is hence in good agreement with predictions for Gaussian fields, even though the temperature fluctuations follow the complex distribution imprinted by gravitational clustering. Being more biased while having a similar variance, the results from the jackknife approach are clearly outperformed by the Gaussian estimate. The results from this section are consistent with our analysis of stochastic fields in Appendix~\ref{sec:toy_models}. In Appendix~\ref{sub:peculiar_velocities}, we furthermore investigate the effect of redshift space distortions on the estimator variance of the angular power spectra and do not find deviations between Gaussian prediction and estimated variance either.

\subsubsection{Correlations}

For a full-sky Gaussian field that is statistically homogeneous and isotropic, we do not expect any correlations between different $\ell$ scales. Non-zero off-diagonal elements in the correlation matrix can however be introduced by the finite mask of the map and through the non-Gaussian nature of the matter density field. Since the PolSpice estimate of the angular power spectrum and its covariance already accounts for the mask, it is particularly interesting to compare the latter with correlation estimates from the independent simulations and the jackknife, thereby enabling us to distinguish masking effects from non-Gaussianity. 

\begin{figure}[t]
	\centering
	\includegraphics[width = .75\linewidth]{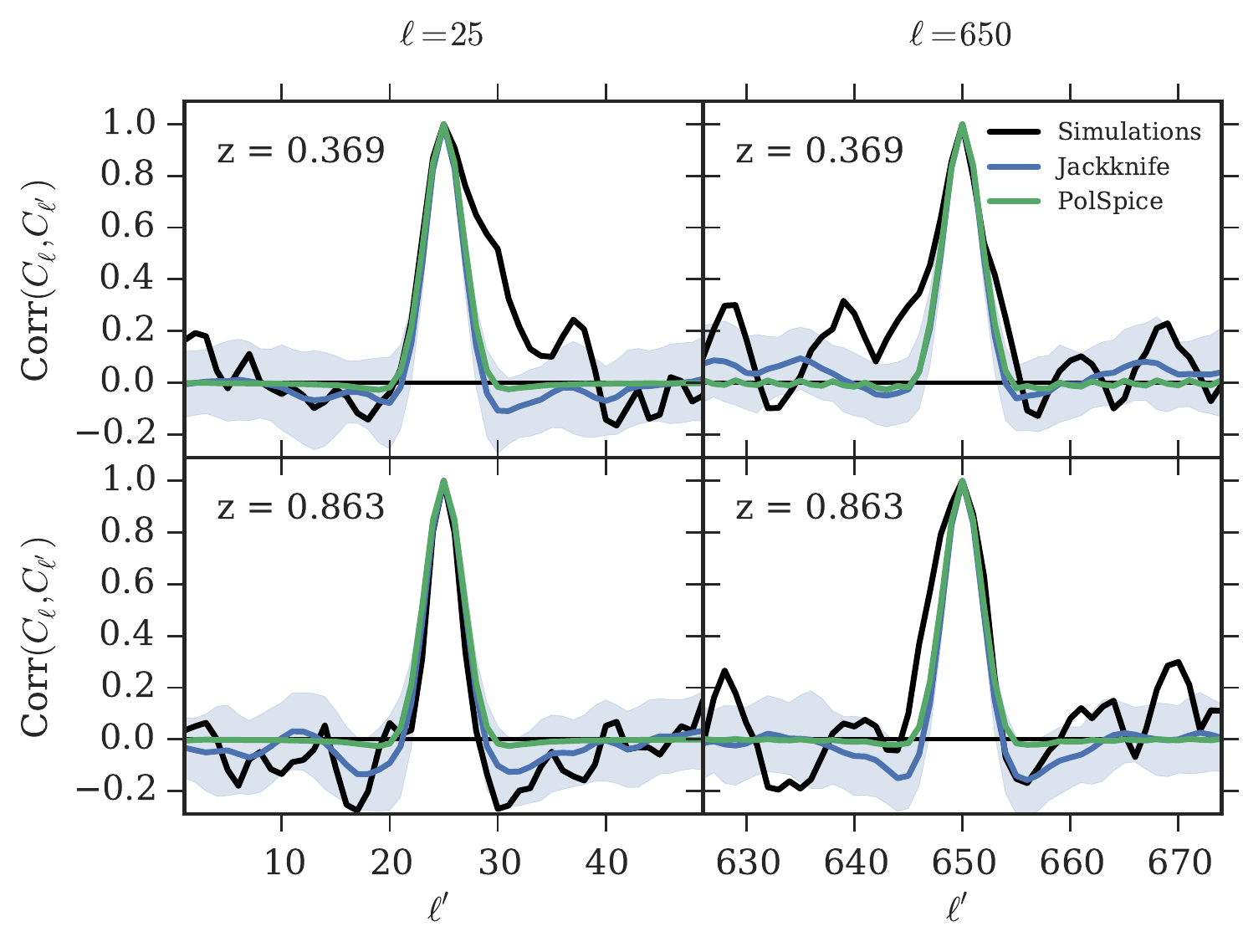}
	\caption{Correlations $\text{Corr}(C_{\ell}, C_{\ell'})$ for the PolSpice angular power spectrum as estimated from the 10 independent simulations, the jackknife, and the Gaussian assumption for the lowest ($z \simeq 0.369$, top panels) and highest ($z \simeq 0.863$, bottom panels) redshift bin of the simulations. Correlations between large scales (around $\ell = 25$) are shown on the left, small scales (around $\ell = 650$) are shown on the right. The shaded areas show the standard deviation of the jackknife estimator from the 10 independent simulations. The estimate from the 10 realizations is consistent with the PolSpice result within the noise.}
	\label{fig:h1_corrs}
\end{figure}

Figure~\ref{fig:h1_corrs} shows $\text{Corr}(C_\ell,C_{\ell'})$ as a function of $\ell'$ for the lowest (top row) and the highest (bottom row) redshift bin and at small ($\ell = 650$, left column) and large scales ($\ell = 25$, right column). Overall, the PolSpice estimate appears to be in good agreement with the simulation estimator within the noise. The jackknife estimate is biased to negative correlations for scales separated by $5 \lesssim \delta \ell \lesssim 15$ and is in good agreement with the simulation and PolSpice estimate otherwise.

\begin{figure}[t]
	\centering
	\includegraphics[width = .75\linewidth]{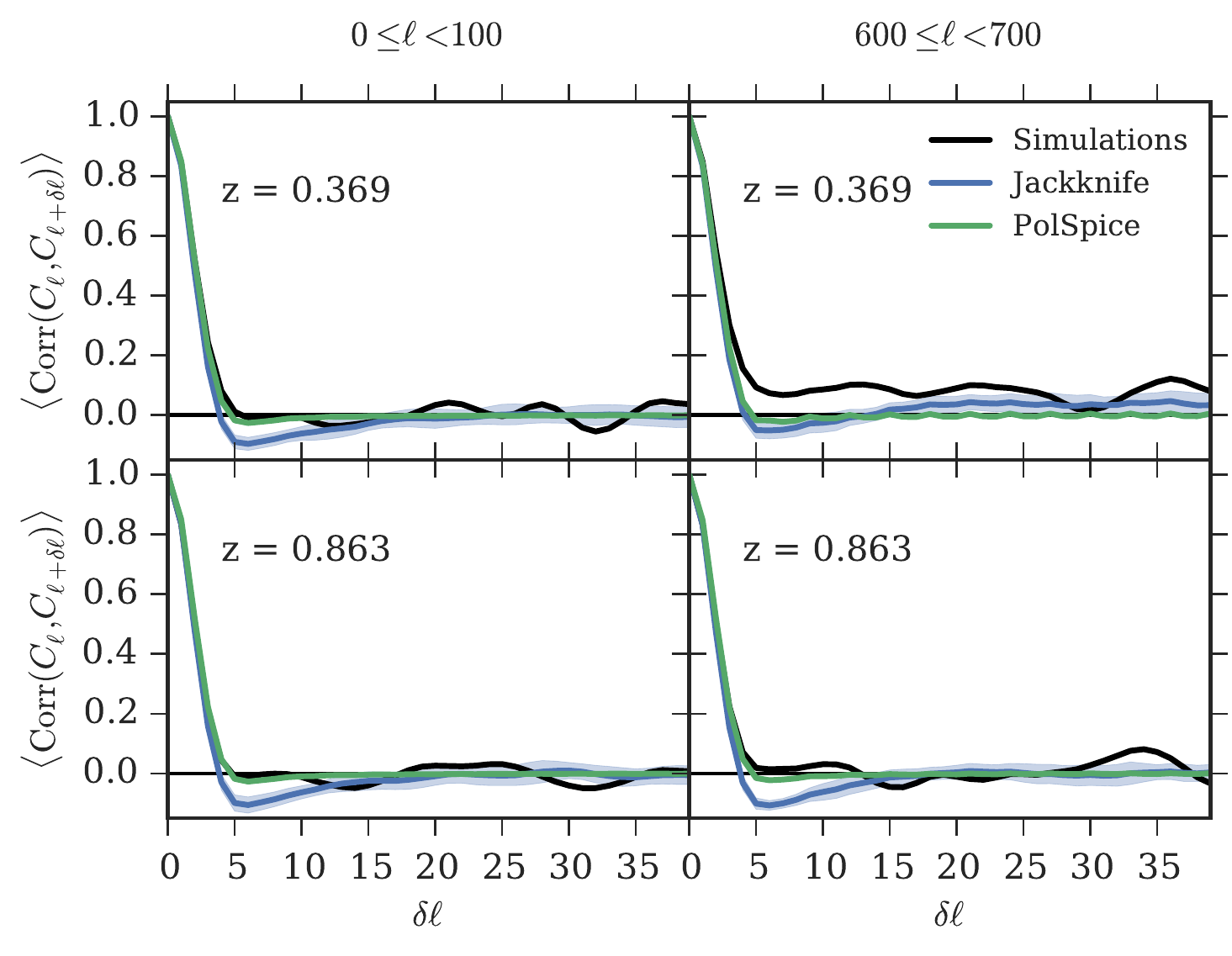}
	\caption{Mean correlation $\langle\text{Corr}(C_{\ell}, C_{\ell+\delta\ell})\rangle$ for the PolSpice angular power spectrum, averaged over 100 $\ell$ values. We show estimates from the independent simulations, the jackknife, and the Gaussian assumption for the lowest ($z \simeq 0.369$, top panels) and highest ($z \simeq 0.863$, bottom panels) redshift bin of our simulations. Correlations between large scales ($0 \leq \ell < 100$) are shown on the left, small scales ($600 \leq \ell < 700$) on the right. The shaded areas show the standard deviation of the jackknife estimator from the 10 realizations.}
	\label{fig:h1_mean_corr}
\end{figure}

To reduce the statistical noise of the simulation estimator, we also show the correlation $\text{Corr}(C_\ell,C_{\ell + \delta \ell})$ as a function of $\delta\ell$ when averaged over neighboring $\ell$ values (we decide to average over 100 $\ell$s). The result is shown in Figure~\ref{fig:h1_mean_corr}. Again, we find good agreement between the PolSpice estimate and the simulation results. Yet, for small scales and low redshifts (top right of Figure~\ref{fig:h1_mean_corr}), there are $\sim 5\%$ extra correlations for $\delta \ell > 15$ in both jackknife and simulation estimators. The bias to negative correlations of the jackknife estimate for $5 \lesssim \delta \ell \lesssim 15$ can be seen in all four panels of Figure \ref{fig:h1_mean_corr}.

The pseudo $C_\ell$ estimator for the covariance neglects correlations, hence its correlation matrix is simply the identity matrix. As expected, both the simulation and the jackknife estimator show that the pseudo $C_\ell$ covariance is not diagonal in practice. Its off-diagonal contributions from mask and non-Gaussianity are very similar to those reported for the PolSpice covariance. 

The overall agreement between the correlations predicted for a masked Gaussian field, the simulation estimator, and the jackknife is remarkably good. While the correlations between individual $\ell$ values estimated from the simulations are too noisy for a detailed comparison, the averaged results in Figure \ref{fig:h1_mean_corr} indicate deviations of approximately $8\%$ for small scales at the lowest redshift $z \simeq 0.369$ which are not detectable at the highest redshift $z \simeq 0.863$. The analysis of Appendix~\ref{sub:peculiar_velocities} shows that this behavior remains unchanged when taking redshift space distortions into account. As for the variance, the results on the correlation matrix are in line with our findings from the analysis of stochastic fields in Appendix \ref{sec:toy_models}. While the correlations introduced through the mask are present at all scales, the correlations from the non-Gaussianity induced by non-linear gravitational clustering only affect small scales. Instrumental noise will also mostly affect small scales, suppressing the contributions from non-Gaussianity for realistic surveys.

\section{Conclusions}
\label{sec:conclusions}

Intensity mapping of the redshifted \SI{21}{cm} emission line of neutral hydrogen (HI) is a promising technique for studying the large-scale distribution of matter in the Universe and will potentially yield complementary information to more traditional galaxy redshift surveys. The majority of the HI content of the low redshift ($z\lesssim1$) Universe is believed to reside in dense clumps inside galaxies. The low angular resolution of radio telescopes combined with high frequency resolution then means that \SI{21}{cm} intensity mapping surveys will map an unresolved (in angular position) population of galaxies containing HI in fine redshift bins. In this paper we have simulated wide field HI intensity maps and used them to study the statistical properties of the expected brightness temperature signal, in particular the angular power spectrum $C_\ell$ and its covariance.

We have simulated HI intensity maps at redshifts $0.35\lesssim z\lesssim 0.9$, based on a suite of 10 N-body simulations of a \SI{2.6}{\per\h\giga\parsec} box~\cite{Wechsler15} and a mass resolution of $\SI{1.6d11}{\per\h\msolar}$. We created HEALPix~\cite{2005ApJ...622..759G} maps of the dark matter density field from the N-body simulations in equally spaced bins in frequency. Since our simulations do not resolve all the low mass halos that are expected to host HI, we have used the coarse density field in each pixel to statistically sample these halos using a halo mass function conditioned on the overdensity of the pixel following~\cite{2002MNRAS.329...61S}. The low angular resolution of the intensity map means that we do not need to assign locations to individual halos within each pixel. Instead, each of these halos is assigned an amount of HI determined by its dark matter mass following a phenomenological prescription based on reference~\cite{Padmanabhan:2015vg}. The resulting distribution of HI has then been used to generate 12 maps of redshifted \SI{21}{cm} brightness temperature with an angular resolution of \SI{7}{\arcmin} between $z \simeq 0.35$ and $z \simeq 0.89$ and redshift bin widths from $\Delta z \simeq 0.033$ at the lowest to $\Delta z \simeq 0.061$ at the highest redshift. 

The main assumptions of our prescription are that the halo distribution within each pixel is determined by its overdensity and well described by the conditional mass function~\cite{2002MNRAS.329...61S}, the linearized overdensity in each pixel is well approximated by the value derived from spherical evolution~\cite{1994ApJ...427...51B}, and that the HI content of the halos is determined by the mass of the halo~\cite{Padmanabhan:2015vg}. As a (non-trivial) validity check for the halo subsampling, we confirmed that our procedure reproduces the unconditional mass function and the large scale bias to better than $10\%$. With high-resolution simulations of smaller volumes, the accuracy of our procedure could be assessed in more detail and we leave such a study to future work. Assessing the validity of the HI assignment is more complicated. The model from~\cite{Padmanabhan:2015vg} is able to reproduce data of sky-averaged values such as the HI density $\Omega_{\rm HI}$ or the bias $b_{\rm HI}$. Due to a lack of data at these redshifts, a more detailed study of the relation between halo and HI on an object by object basis would have to resort to hydrodynamic simulations (see~\cite{2016MNRAS.456.3553V} for example). We furthermore showed a way for including redshift space distortions in our simulations and verified that they do not affect the main conclusions drawn from the analysis of the undistorted maps in the following.

Our brightness temperature maps of the cosmic signal are sufficiently realistic to study the impact of non-linear effects on the large-scale clustering properties of HI intensity maps. In our analysis of the maps, we have focused on the angular clustering of the brightness temperature fluctuations as measured by the angular power spectrum. We have used two estimators for the power spectrum: the pseudo $C_\ell$ approach~\cite{1973ApJ...185..413P} and the publicly available PolSpice package~\citep{2001ApJ...561L..11S, Chon:2004ki}. We have verified that the estimated angular power spectra of the simulated intensity maps agree well with predictions from linear theory and the halo model on large scales. We have found that the bias of HI relative to dark matter, as estimated from the ratio of angular power spectra, is close to unity and mildly scale dependent. For the lowest redshift of our simulations ($z \approx 0.366$), the bias falls below its large scale value by $\sim 10\%$ at $\ell \approx 200$.

Using the multiple N-body realizations, we have estimated the covariance of both $C_\ell$ estimators and compared the results to estimators from jackknife resampling and analytic predictions based on Gaussian statistics. Treating the covariance from multiple simulations as a noisy estimate of the true covariance, we have found good agreement with the PolSpice covariance estimate based on Gaussian statistics. This shows that even for our simple survey geometry (a contiguous quarter of the sky), most off-diagonal correlations are introduced by the mask. Only for small scales and low redshifts, the results indicate an excess of $\sim 8\%$ off-diagonal non-Gaussian correlations that are not captured by the PolSpice estimate. The jackknife estimator overpredicted the error on the angular power spectrum by approximately $\sim 40\%$, independently of scale, redshift and $C_\ell$ estimator. It was however able to trace the extra correlations introduced by the non-Gaussian nature of the brightness temperature fluctuations. The pseudo $C_\ell$ estimator for the covariance has no off-diagonal correlations by construction and overestimated the variance by a factor of two. Looking at the variance of pseudo $C_\ell$ bandpowers, i.e. binned $C_\ell$ values that are approximately uncorrelated, the lack of correlations however balanced the excess in variance and the estimate converged to the results from the multiple realizations. It is worth noting that the HI prescription from~\cite{Padmanabhan:2015vg} leads to a distribution of brightness temperature fluctuations that is less skewed to high contrasts than the matter density field and hence also more Gaussian. The lack of HI in high overdensities is supported by physical~\cite{Bagla:2010fu} and observational~\cite{2011AJ....142..170H,2013ApJ...776...43P} arguments, but changes in the HI prescription could nevertheless influence some of the conclusions on the covariance of the angular power spectrum drawn from our maps.

Our analysis showed that our approach to simulating HI intensity maps can be used to simulate and study upcoming surveys. It also allows for the development of data analysis techniques that improve the extraction of cosmological information and the separation of the cosmological signal and contaminating components such as astrophysical foregrounds or human-made radio frequency interference. As galaxy surveys cover greater parts of the sky to increasing depth, cross-correlations between intensity mapping and galaxy surveys are of particular interest as they are less sensitive to systematic effects. In future work, such simulations can thus be used to additionally model the connection between halos and galaxies in order to study the cross-correlation signal between HI and galaxy surveys.

\acknowledgments
We thank Matthew Becker and Risa Wechsler for providing the cosmological simulations used in this work. The simulations were run using computing resources at the University of Chicago and NERSC. We also thank Andrina~Nicola, Hamsa~Padmanabhan, Oliver~Hahn, Tirth~Choudhury, and Raghunathan~Srianand for insightful discussions, Francesco~Montanari for his support with CLASS, and an anonymous referee for useful comments that helped improve the manuscript. Some of the results in this paper have been derived using the HEALPix~\cite{2005ApJ...622..759G} package. This work was in part supported by the Swiss National Science Foundation (Grant No. $200021\_143906$).

\bibliographystyle{JHEP}
\bibliography{imSims}

\appendix

\section{Sub-grid halo sampling}
\label{sec:sub_grid}

In this Appendix, we compare our prescription for sampling halos from the density field with the resolved halos in the simulation. Our sub-grid model involves the linearization of the non-linear density field~\cite{1994ApJ...427...51B} and the sampling of halos from the conditional mass function from~\cite{2002MNRAS.329...61S}. In the following, we first analyze the linearization before proceeding to a comparison between the resolved and sub-sampled halo populations.

\subsection{Linearization}
\label{sub:linearization}

\begin{figure}[t]
	\centering
	\includegraphics[width = .6\linewidth]{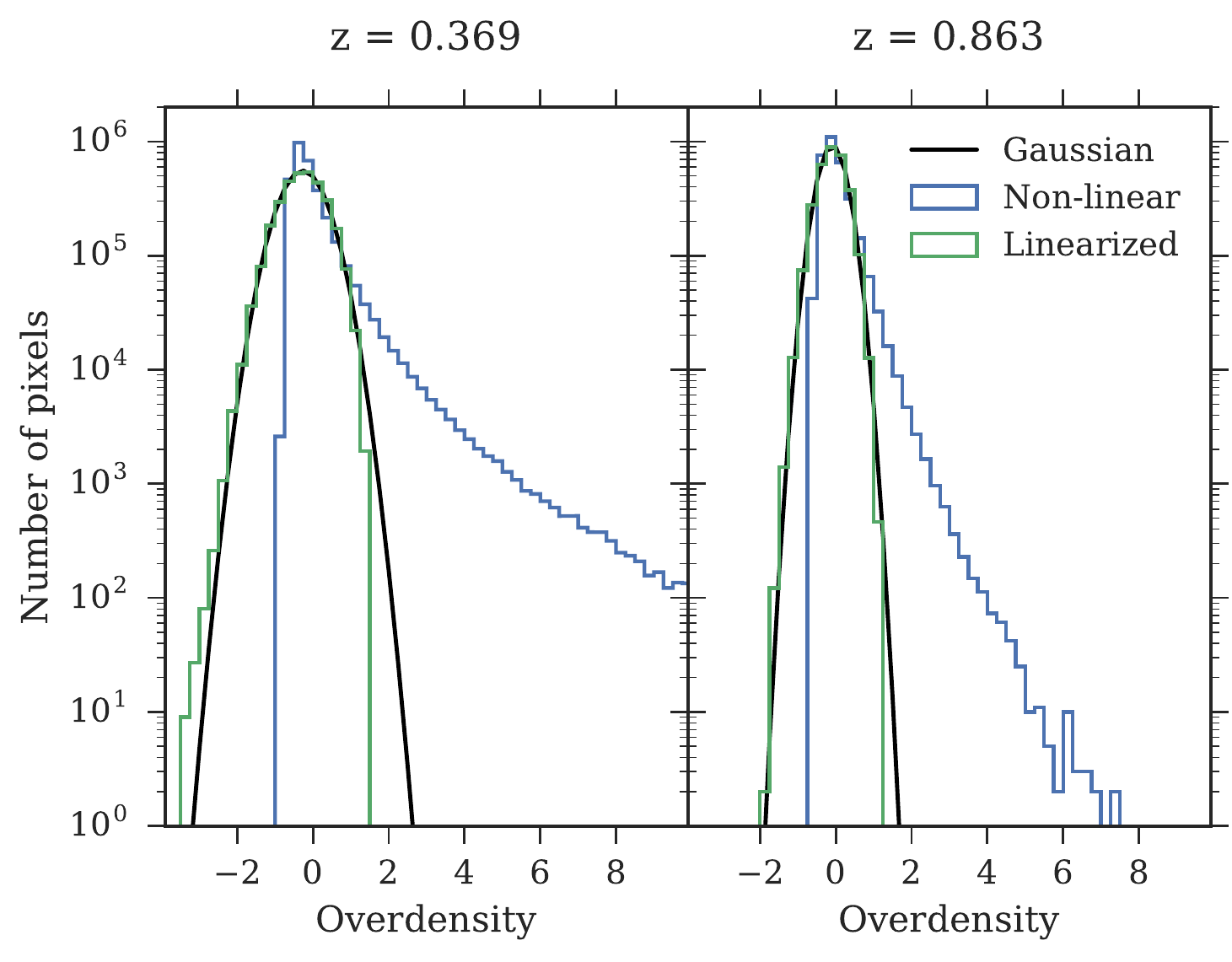}
	\caption{Histogram of non-linear (blue) and linearized (green) matter overdensities in our maps at the lowest ($z \simeq 0.369$, left) and highest ($z \simeq 0.863$, right) redshift. The non-linear overdensities have a tail towards high overdensities and are, by definition, cut off at an underdensity of -1. The distribution of the linearized overdensities is overplotted with a Gaussian distribution of the same mean and variance (black line). We can see that the linearized overdensities approximately follow the Gaussian distribution, but are shifted and skewed to negative values in particular for the lowest redshift bin.}
	\label{fig:linearized_deltas}
\end{figure}

Figure~\ref{fig:linearized_deltas} shows the distribution of overdensities in the dark matter maps before and after linearization for the lowest and highest redshift bin. We expect the linearized overdensities to follow a Gaussian distribution of mean zero, but the variance of the Gaussian depends on the filtering of the field by our pixels and is hence hard to predict from the linear power spectrum. We therefore also plot a Gaussian with the same mean and variance as the distribution of linearized overdensities. We can see in Figure \ref{fig:linearized_deltas} that the agreement is reasonable, but the distribution of overdensities is shifted (means of $-0.27$ and $-0.09$ for the low and high redshift bin) and skewed (skewness of $-0.14$ and $-0.10$ for the low and high redshift bin) towards negative overdensities in particular for the lowest redshift bins. This means that equation~\eqref{eq:deltalin} fails at correctly reproducing the high overdensity tail of the linearized field for our pixels.

\begin{figure}[t]
	\centering
	\includegraphics[width = .6\linewidth]{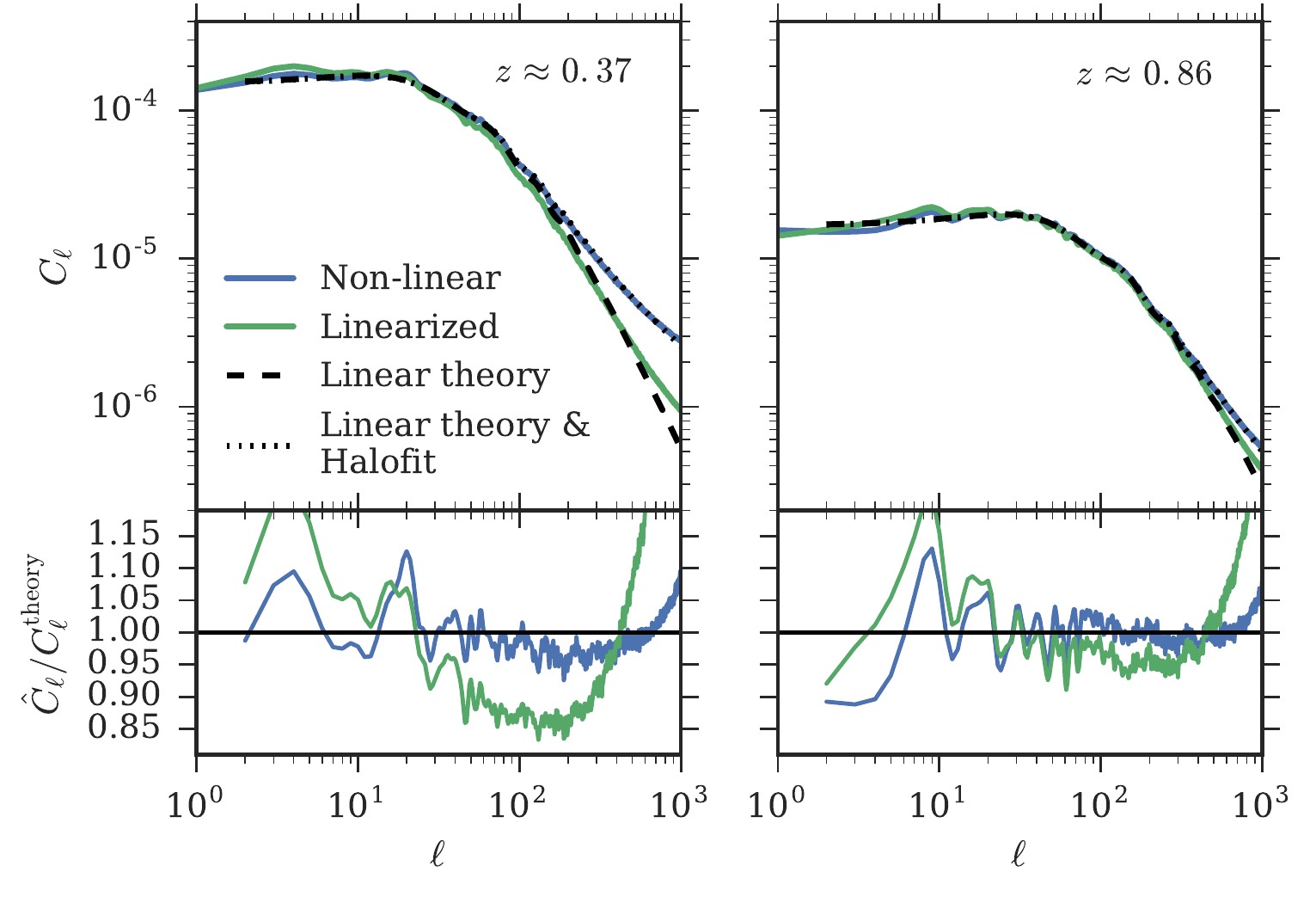}
	\caption{Angular power spectrum of non-linear (blue) and linearized (green) matter overdensities. The spectrum of the linearized field and is supposed to follow the linear theory power spectrum (black dashed line). We can see that, in particular for the lowest redshift bin, the estimated and predicted linear power spectra deviate by around $15\%$.}
	\label{fig:linearized_cells}
\end{figure}

Figure \ref{fig:linearized_cells} shows the angular power spectrum of the non-linear and linearized field as compared to the respective predictions from linear theory and halofit corrections. We can see that, overall, the correlations of the linearized field follow the linear theory prediction. The bottom panel shows, however, that the linearized field has deviations of around $15\%$ when compared to its prediction while the non-linear field agrees with the non-linear prediction to better than $5\%$.

\subsection{Comparison of halo populations}
\label{sub:comparison_of_halo_populations}

In this section, we compare the \emph{resolved halos} in the simulation to the halos sampled from the density field of the same simulation with our procedure. We used rockstar~\cite{2013ApJ...762..109B} for finding halos in the lightcones of the simulation and use the mass $M_{200, c}$ within the region where the density is 200 times higher than the critical density as halo mass.

\begin{figure}[t]
	\centering
	\includegraphics[width = .6\linewidth]{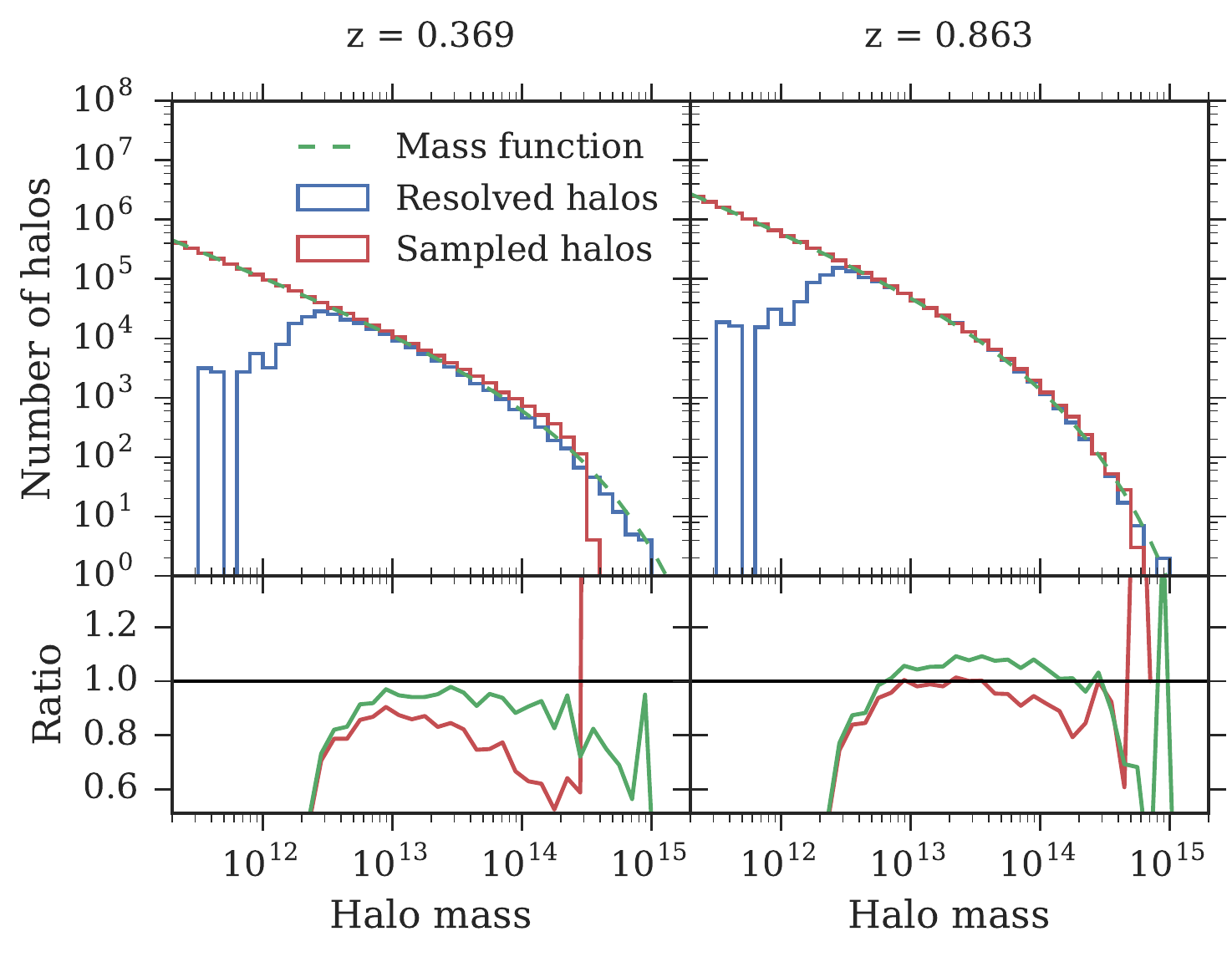}
	\caption{Comparison of the number of halos within a given mass bin from the simulation (blue histograms) and the sampled halos (red histogram) in the lowest (left) and highest (right) redshift bin. The green line shows the prediction from the unconditional mass function~\cite{2002MNRAS.329...61S}. We can see that the halo catalog from the simulation is complete for halos with mass $M_{200,c} > 10^{13}\si{\per\h\msolar}$. The distribution of the sampled halos agrees with the resolved halos for masses close to this limit, but deviates for the highest masses in particular for the low redshift bin.}
	\label{fig:mf_comp}
\end{figure}

Figure \ref{fig:mf_comp} shows the agreement between the number of halos within a given mass range as observed from the population of resolved and sampled halos as well as the distribution predicted from the unconditional mass function from reference~~\cite{2002MNRAS.329...61S}. The simulations have a mass resolution of $\SI{1.6d11}{\per\h\msolar}$ and the distribution of resolved halos is therefore not complete for masses below approximately $10^{13}\si{\per\h\msolar}$. The exponential cut-off scale in the HI assignment, however, is at roughly $10^{12}\si{\per\h\msolar}$ so in our model the resolved halos would not contribute significantly to the overall HI distribution. Close to $10^{13}\si{\per\h\msolar}$, the resolved and sampled halo populations indeed agree to about $10\%$. The high mass tail of the sampled halo distribution at low redshifts, however, diverges from the resolved population. This is most likely due to the shortcomings of the linearization process at the high overdensities discussed in section~\ref{sub:linearization}. As these halos do not contain significant amounts of HI in our prescription, this mismatch is however not critical for our application. At the highest redshift bin our method seems to extend even to this high mass end.

\begin{figure}[t]
	\centering
	\includegraphics[width = .6\linewidth]{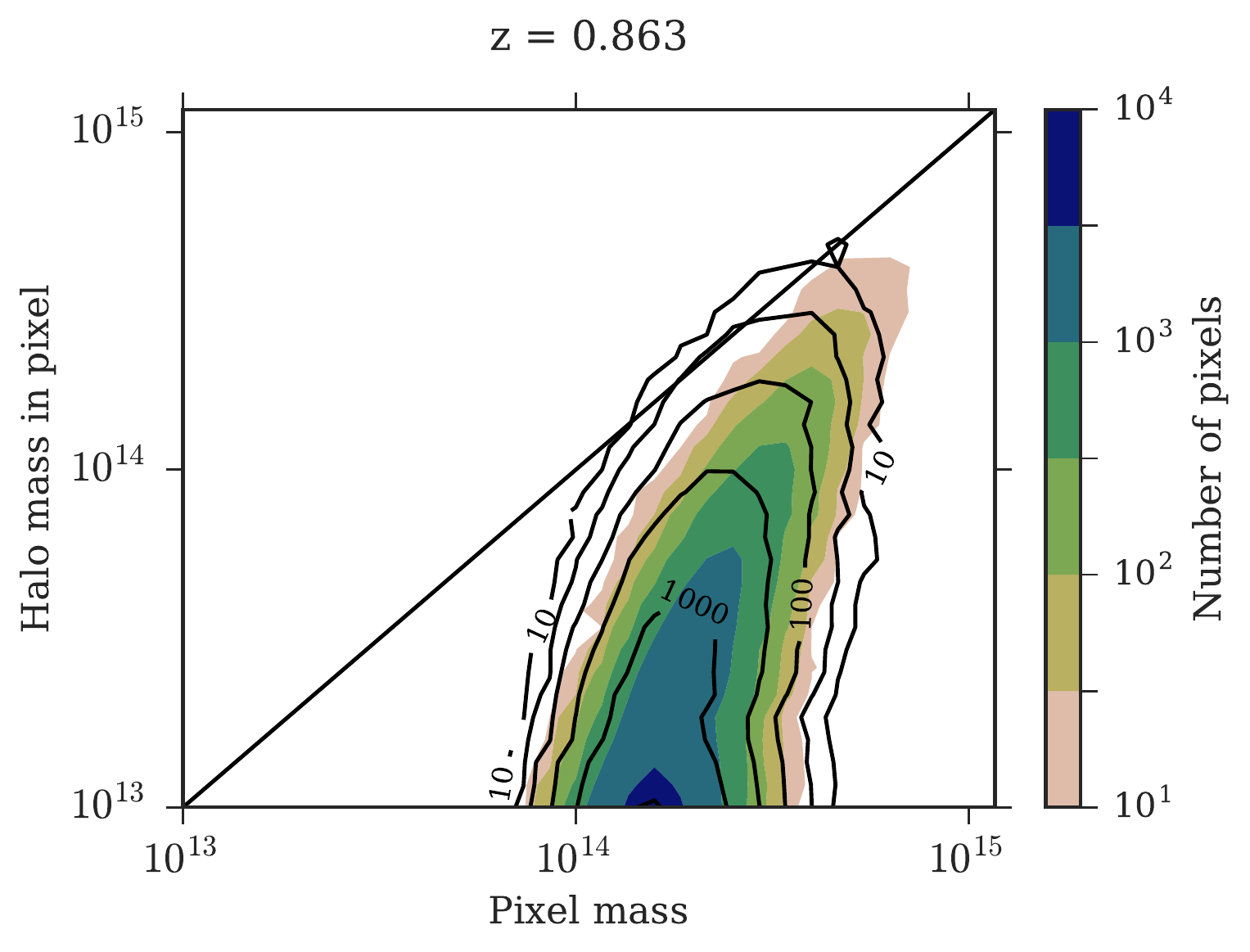}
	\caption{Comparison of mass distributions in pixels for the density field and the halos in the highest redshift bin. We only consider halos with mass greater than $10^{13}\si{\per\h\msolar}$. The filled contours show the distribution of the resolved halos and the line contours show the distributions of the sampled halos. Note that the plot is not showing the pixels which contain no halos.}
	\label{fig:pixel_vs_halomass}
\end{figure}

In Figure \ref{fig:pixel_vs_halomass} we show the distribution of pixel masses in the matter density and the halo field at the highest redshift. The halos are either the resolved halos from the simulations (filled contours) or the sampled halos (lines). The plot shows that both the distribution and the scatter in the halo mass distribution of the sampled halos is very similar to the distribution of the resolved halos. There are mild differences in the scatter that are most likely related to the simple Poisson sampling scheme. For the same reason, some of the pixels slightly violate mass conservation. None of these shortcomings are expected to affect the halos within the mass range relevant for our intensity maps.

\subsection{Peculiar velocities}
\label{sub:peculiar_velocities}

When assigning the halos to pixels, one has to take into account that, depending on the line-of-sight velocities of the halos and the position of the halos within the radial extension of the bin, the \SI{21}{cm} line in some of the halos will shift into the neighboring frequency bins. The bin-width of our maps corresponds to a frequency bandwidth of $\Delta f = \SI{25}{\mega\hertz}$. For a halo at the center of the bin to escape the bin, it would thus require velocities larger than $\SI{3d3}{\kilo\meter\per\second}$ which are rarely achieved. At the boundaries of the bin, however, the redshift space distortions (RSDs) due to peculiar velocities are distorting the shape of this boundary.

\begin{figure}[t]
	\centering
	\includegraphics[width = .8\linewidth]{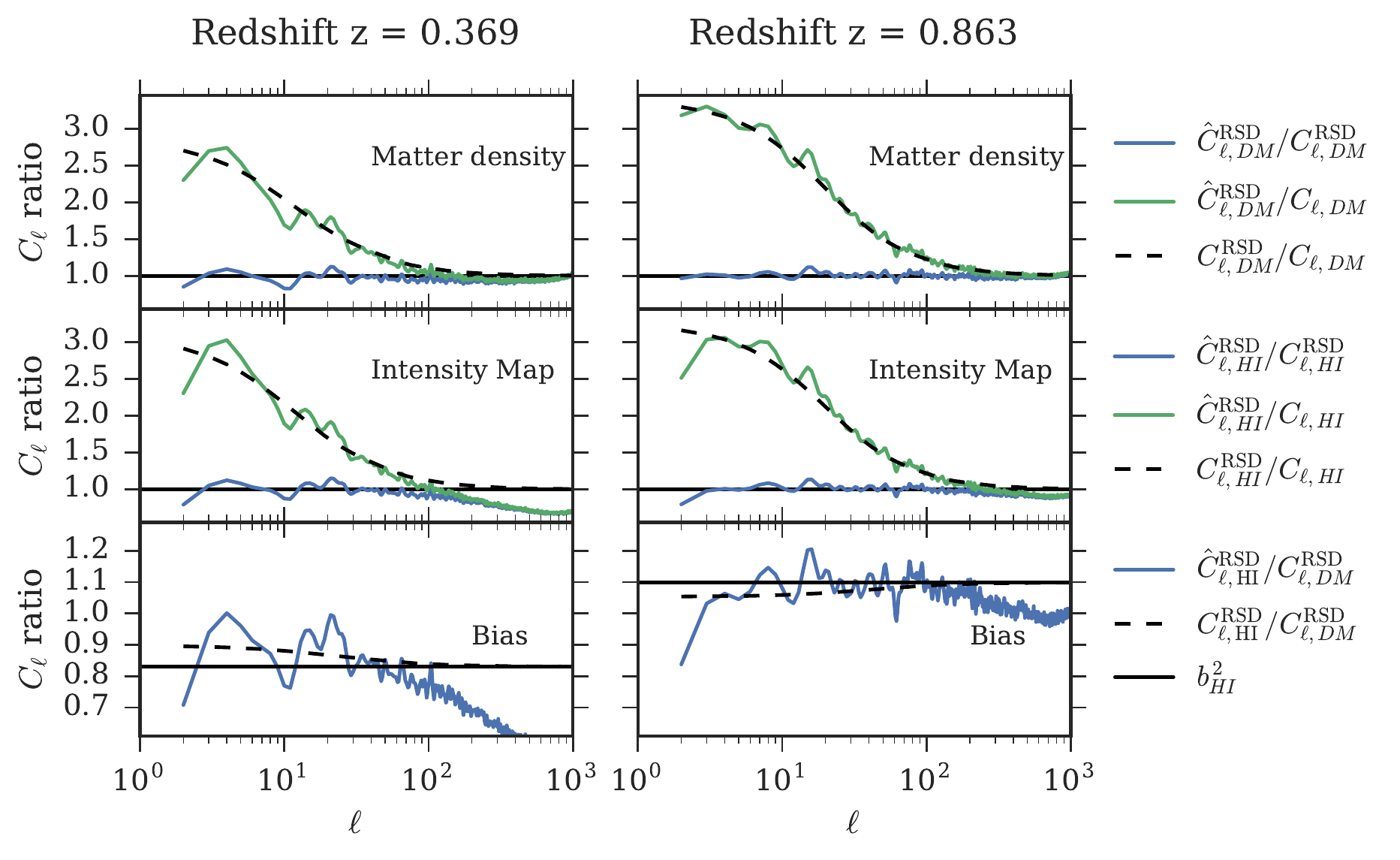}
	\caption{Comparison of the angular power spectra of the redshift space distorted maps with theoretical predictions with (blue) and without (green) redshift space distortions for the lowest (left) and highest (right) redshift bins. The dashed black line shows the ratio of both theoretical predictions. The top panels show the results for the matter density field and the middle panels show the results for the intensity maps. The prediction for the intensity maps assume a scale independent bias that does not evolve within the redshift bins and is derived from the unconditional mass function in~\cite{2002MNRAS.329...61S}. The bottom panel shows the ratio of the spectrum estimated from the intensity maps and the spectrum predicted for dark matter, i.e. it shows the biasing of the intensity maps with respect to the matter maps. The solid line shows the predicted large scale bias and the dashed line shows the ratio between the theoretical prediction for the intensity maps and the matter maps.}
	\label{fig:rsd_cells_ratios}
\end{figure}

To model the effect of RSDs on our maps, we start, as before, with dark matter maps for which we assign each particle of the N-body simulation within the lightcone to the bin which corresponds to its position in redshift according to its comoving distance. In addition to the number $n_{ij}$ of particles within pixel $i$ of redshift bin $j$, we now additionally count the number $\tilde n_{ij}(k)$ of particles within this volume that fall into redshift bin $k$ when furthermore taking their line of sight velocity into account. The number of particles $\tilde n_{ik}$ within pixel $i$ of redshift bin $k$ of the distorted dark matter maps (including RSDs) is then given by $\tilde n_{ik} = \sum_j \tilde n_{ij}(k)$. Only a small fraction of the particles gets redistributed by this procedure, but due to the coherence of the effect on large scales, even these mild changes enhance the large scale power spectrum of the field~\cite{1987MNRAS.227....1K}. We created the distorted dark matter maps from four independent simulations and Figure~\ref{fig:mattervssimcells} shows a comparison of the angular power spectra as estimated from those maps with (thin green line) and without (thick green line) RSDs. The top panel of Figure \ref{fig:rsd_cells_ratios} furthermore shows that the enhancement of the large scale correlations in the distorted maps is as expected from the theoretical prediction by CLASS (see~\cite{Dio:2013ia} for details on RSDs in CLASS).

Our approach for assigning halos to pixels is based on the insight that the number of halos within each pixel can be modeled using the conditional mass function of reference~\cite{2002MNRAS.329...61S}. The procedure however does not model the positions of the halos within the pixels or their velocities. Assuming that the halo velocities are unbiased with respect to the dark matter flow (see for example~\cite{2010PhRvD..81b3526D} for a discussion), we can however use the information about the distorted dark matter particle distribution $\tilde n_{ij}(k)$ in order to imprint the RSDs on the HI intensity maps. Starting from the unperturbed dark matter density field, we follow the procedure described in section~\ref{sec:simulating_h_i_intensity_maps} to assign an HI mass $m^{\rm HI}_{ij}$ to pixel $i$ of redshift bin $j$. Instead of adding the complete HI mass $m^{\rm HI}_{ij}$ to this pixel, we however distribute the mass over all redshifts according to the fraction $\tilde n_{ij}(k)/n_{ij}$ of dark matter particles that got redistributed due to their line of sight velocity. In pixel $i$ of redshift bin $k$, the distorted HI maps consequently contain a mass of 
\begin{equation}
	\tilde m^{\rm HI}_{ik} = \sum_j m^{\rm HI}_{ij} \tilde n_{ij}(k)/n_{ij}.
\end{equation}

Figure \ref{fig:mattervssimcells} shows the angular power spectrum of the intensity maps with (thin blue line) and without (thick blue line) RSDs. As expected, the power on large scales is enhanced. In the middle panel of Figure \ref{fig:rsd_cells_ratios} it can be seen that the power spectrum follows the theoretical prediction for a biased tracer with unbiased velocities. The bottom panel furthermore shows the agreement of the large scale bias with the analytical predictions.

\begin{figure}[t]
	\centering
	\includegraphics[width = .46\linewidth]{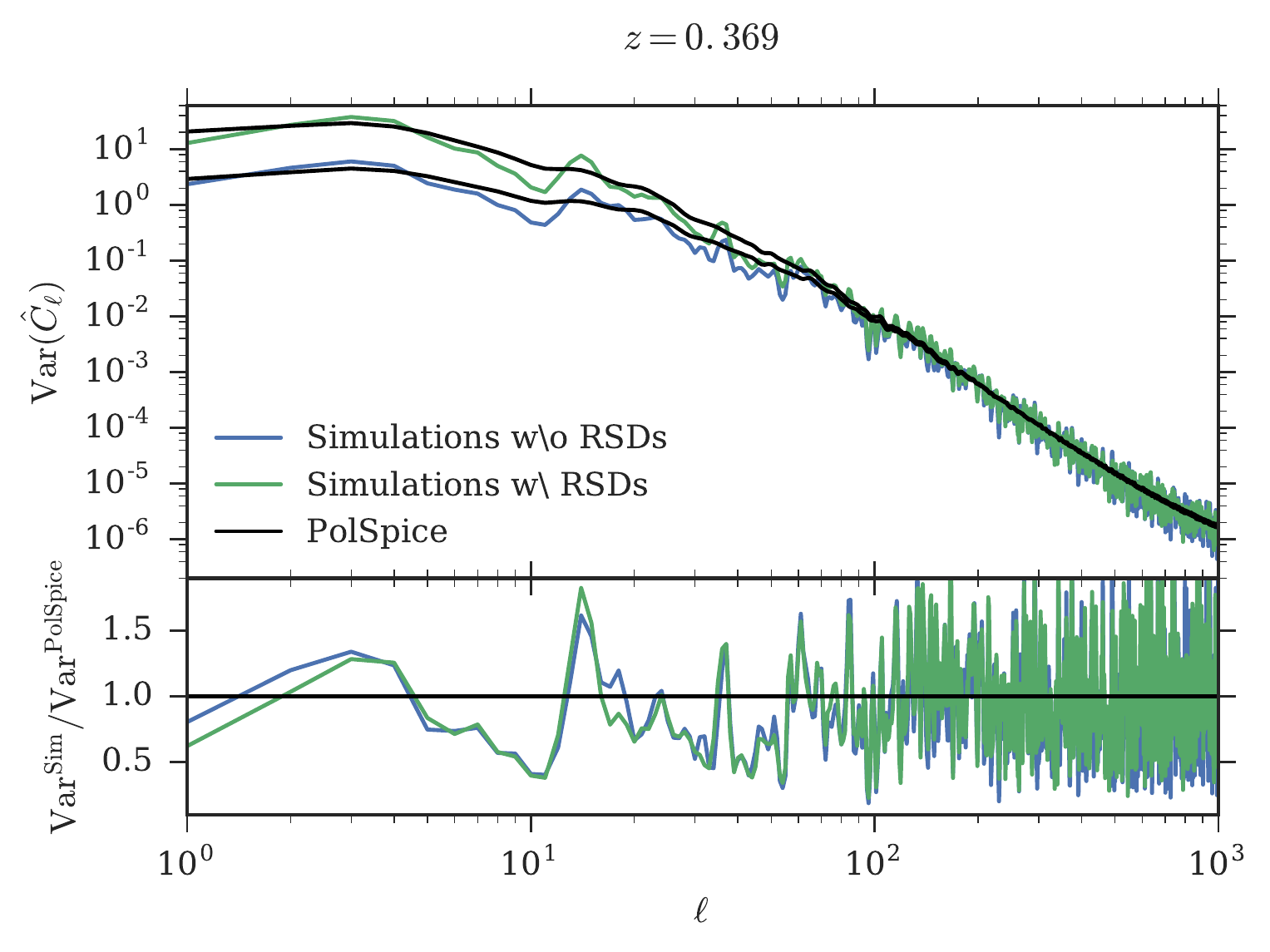}
	\includegraphics[width = .46\linewidth]{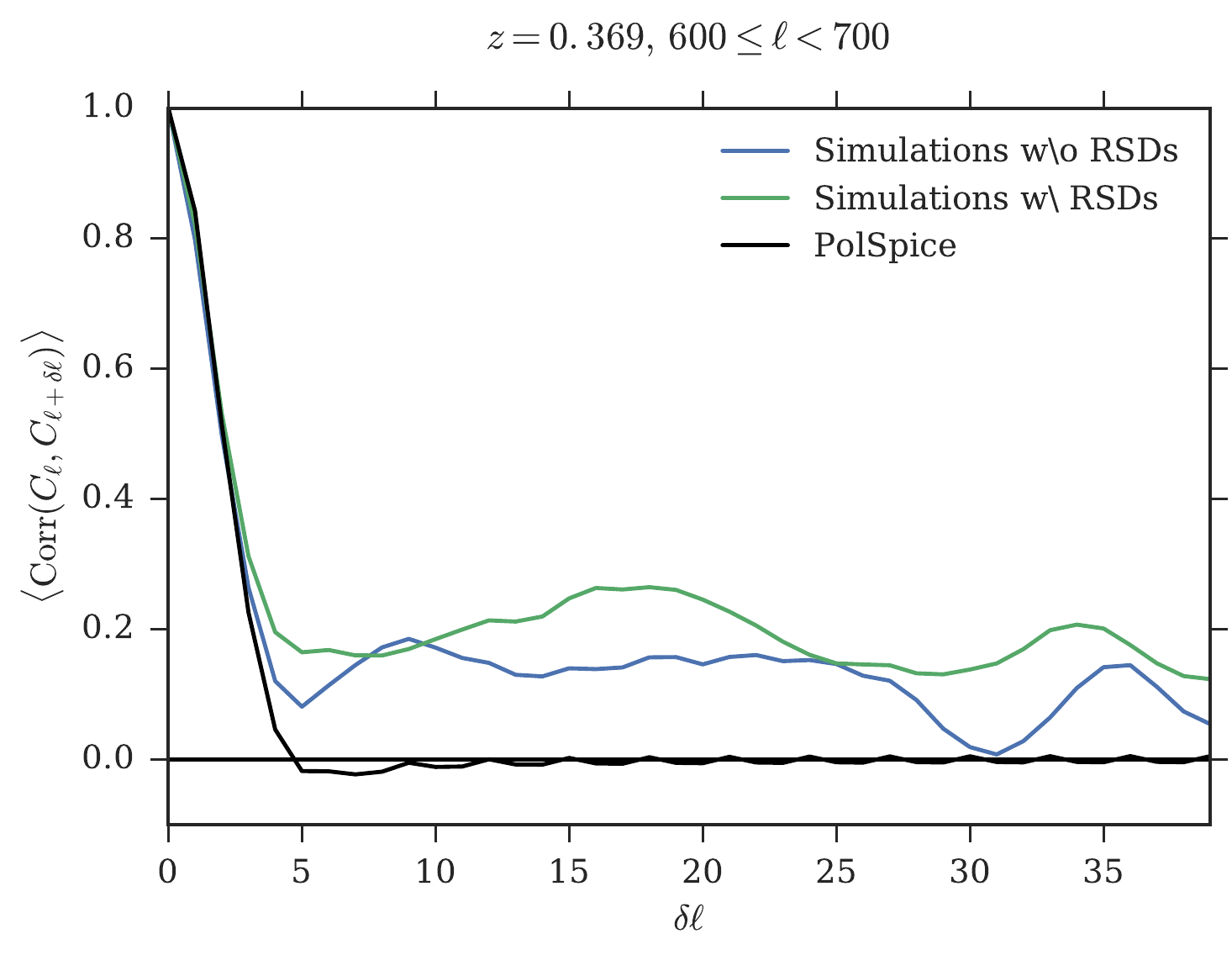}
	\caption{Variance (left) and correlations (right) of the angular power spectra of the dark matter field with (green) and without (blue) RSDs as estimated from our analysis of four realizations of the simulations. The corresponding estimates for Gaussian random fields from PolSpice are shown in black. The bottom panel of the variance plot shows the ratio of the estimate from the simulations to the PolSpice predictions. To reduce the noise in the correlation matrix estimate, we show the correlations between neighboring $\ell$ values averaged over $600 \leq \ell < 700$.}
	\label{fig:dm_errs_rsd}
\end{figure}

Using the intensity maps from four realizations of the N-body simulation, we furthermore studied if the RSDs affect the agreement between the Gaussian estimate and the estimate from the independent realizations for variance and covariance of the HI angular power spectra. Figure \ref{fig:dm_errs_rsd} shows the results for the variance ${\rm Var}(\hat C_\ell)$ and averaged correlations $\langle{\rm Corr}(\hat C_\ell, \hat C_{\ell + \delta \ell})\rangle$ between small scales ($600 \leq \ell < 700$, see also section \ref{sub:results_cov} for more discussion of the averaged correlation). The left plot shows that, within the noise of the estimators, the variance follows the Gaussian prediction from PolSpice for both the simulations with and without RSDs. We verified that the same is true for the correlation matrix at high redshifts or large scales. As for the simulations without RSDs, we find an excess of correlations between neighboring $\ell$ scales for high $\ell$ values ($600 \leq \ell < 700$). The right plot of Figure~\ref{fig:dm_errs_rsd} however shows that, again within the noise of the estimate, the magnitude of the excess is broadly consistent with the estimate from the maps without RSDs.

\section{Stochastic field models}
\label{sec:toy_models}

In this section we analyze the properties of pseudo $C_\ell$ and PolSpice estimators when applied to masked random fields. We will compare the results from many realizations of maps with the same power spectrum to the results from both Gaussian field predictions and resampling methods. The mask is chosen such that it matches the mask which we also impose on our simulated intensity maps. We will use two types of random fields: Gaussian fields and log-normal fields.

\subsection{Gaussian fields}
\label{sub:gaussian_fields}

As input power spectrum, we use the angular power spectrum for redshift $z \simeq 0.863$ with a bin width of $\Delta z \simeq 0.06$ as calculated from linear theory with halofit corrections using CLASS~\citep{Blas:2011jb, Dio:2013ia}. The highest redshift map is closest to a Gaussian random field and closest to the linear theory prediction, so this analysis is used as a mock for the high redshift behavior of the covariance. We rescale each density field to an intensity map using equation~\eqref{eq:naiveT}. To estimate the variance and covariance of the angular power spectrum estimator, we apply jackknife and the predictions for Gaussian fields to $100$ random realizations of the field. We compare these results to the values inferred from $10^4$ random realizations from the HEALPix routine \verb|synfast|.

\begin{figure}[t]
	\centering
	\includegraphics[width = .49\linewidth]{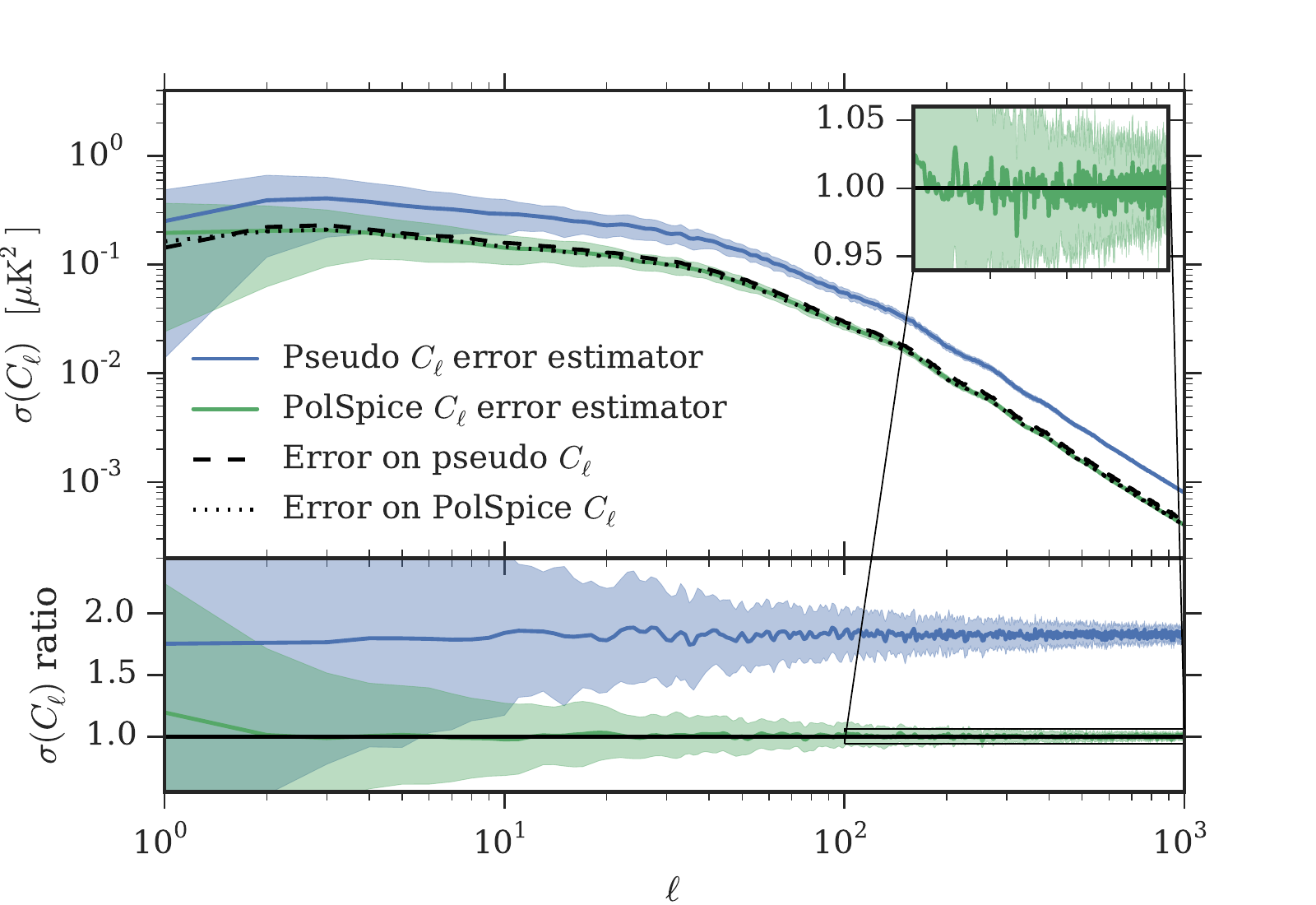}
	\includegraphics[width = .49\linewidth]{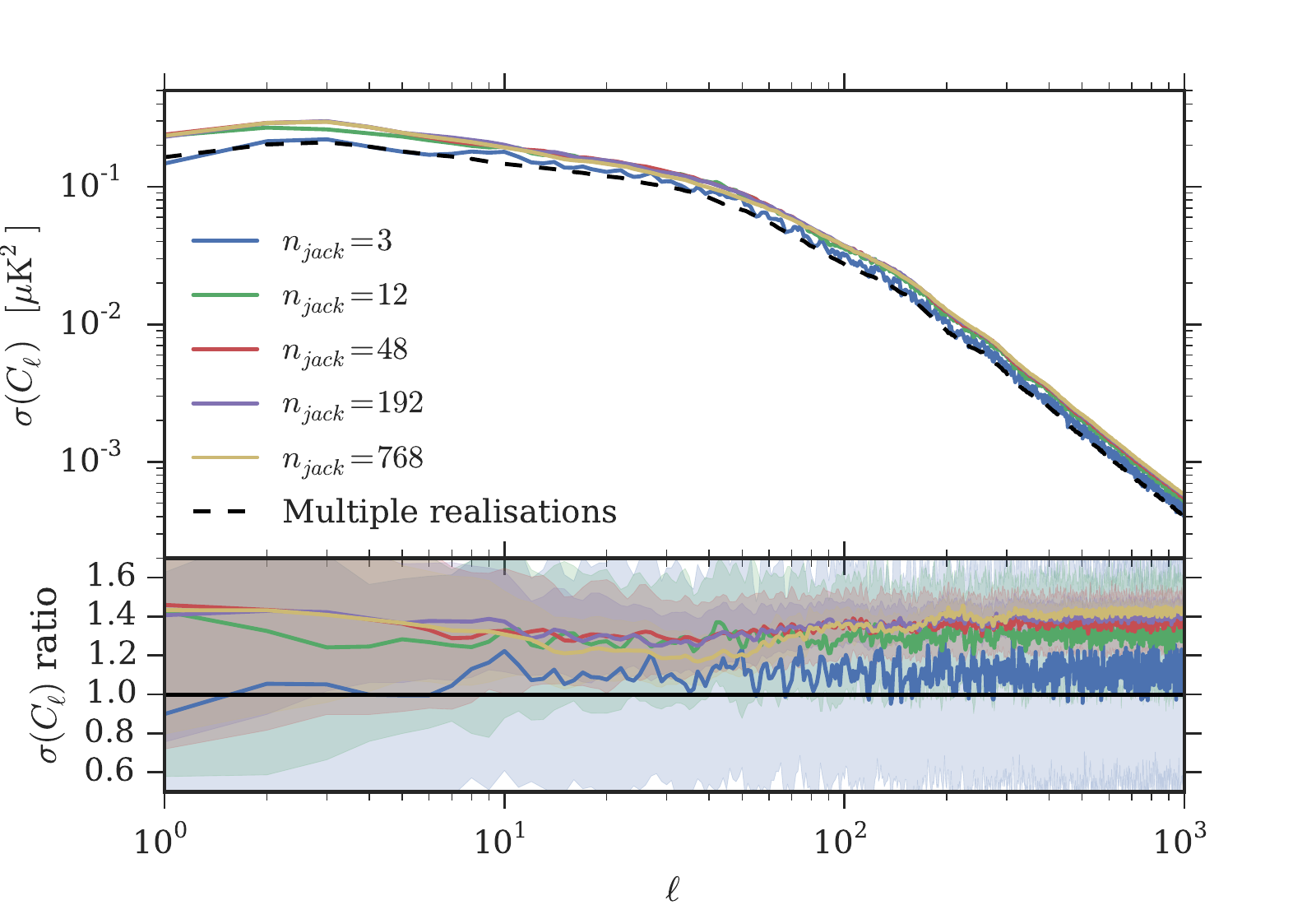}
	\caption{Standard deviation $\sigma(C_\ell)$ of the angular power spectrum from multiple realizations of the masked Gaussian field (black) as compared to the pseudo $C_\ell$ (blue) and PolSpice (green) estimators for Gaussian fields (left panel) and to different jackknife configurations (right panel). While the PolSpice estimate is consistent with the result from $10^4$ realizations, the pseudo $C_\ell$ estimate overestimates $\sigma(C_\ell)$ by a factor of $\sim 2$ to compensate for the lack of correlations. As demonstrated in section \ref{sub:results_cov} for the simulated intensity maps, this mismatch can be reduced by considering bandpowers that are binned in $\ell$. The jackknife error estimates become more and more biased to higher values as $n_{\rm jack}$ increases, while the variance of the error estimate decreases.}
	\label{fig:gfcellerrors}
\end{figure}

Considering the diagonal of the covariance matrix $\sigma^2(C_\ell) = \text{Cov}(C_{\ell}, C_{\ell})$, the left panel of Figure~\ref{fig:gfcellerrors} shows a comparison of $\sigma(C_\ell)$ estimates from Gaussian fields together with the results from the $10^4$ realizations of a Gaussian random field with the same input power spectrum. The pseudo $C_\ell$ estimator overestimates $\sigma(C_\ell)$ by a factor of $\sim 2$, while the prediction from PolSpice agrees well with the result from the $10^4$ realizations. The pseudo $C_\ell$ estimate for the error is getting closer to the correct value if we consider the variance of bandpowers that are averaged over bins in $\ell$. This however comes at the cost of loosing more and more information on the spectrum itself due to the averaging.

The right panel of Figure~\ref{fig:gfcellerrors} shows the results for the jackknife variance estimates. We use the PolSpice estimator for estimating the angular power spectrum of the maps, but the results are similar when using pseudo $C_\ell$ values. To create the jackknife sample, we mask patches that correspond to pixels in coarser HEALPix maps. Each color in the right panel of Figure~\ref{fig:gfcellerrors} refers to a different patch size. The number of jackknives $n_{\rm jack}$ given in the legend is inversely proportional to the size of the patch. We can see that the error estimate with the smallest bias is coming from the jackknife with largest patch size. The higher we go in $n_{\rm jack}$---i.e. the lower in patch size---the smaller is the noise on the estimate but the larger is the disagreement between the jackknife estimate and the repeated simulations. To compromise between variance and bias of the jackknife estimate, we chose $n_{\rm jack} = 48$ as our jackknife configuration for the remaining analysis. For a jackknife with $n_{\rm jack} = 48$, $\sigma(C_\ell)$ is biased to higher values by about $40\%$.

\subsection{Log-normal fields}
\label{sub:lognormal_fields}

\begin{figure}[t]
	\centering
	\includegraphics[width = .6\linewidth]{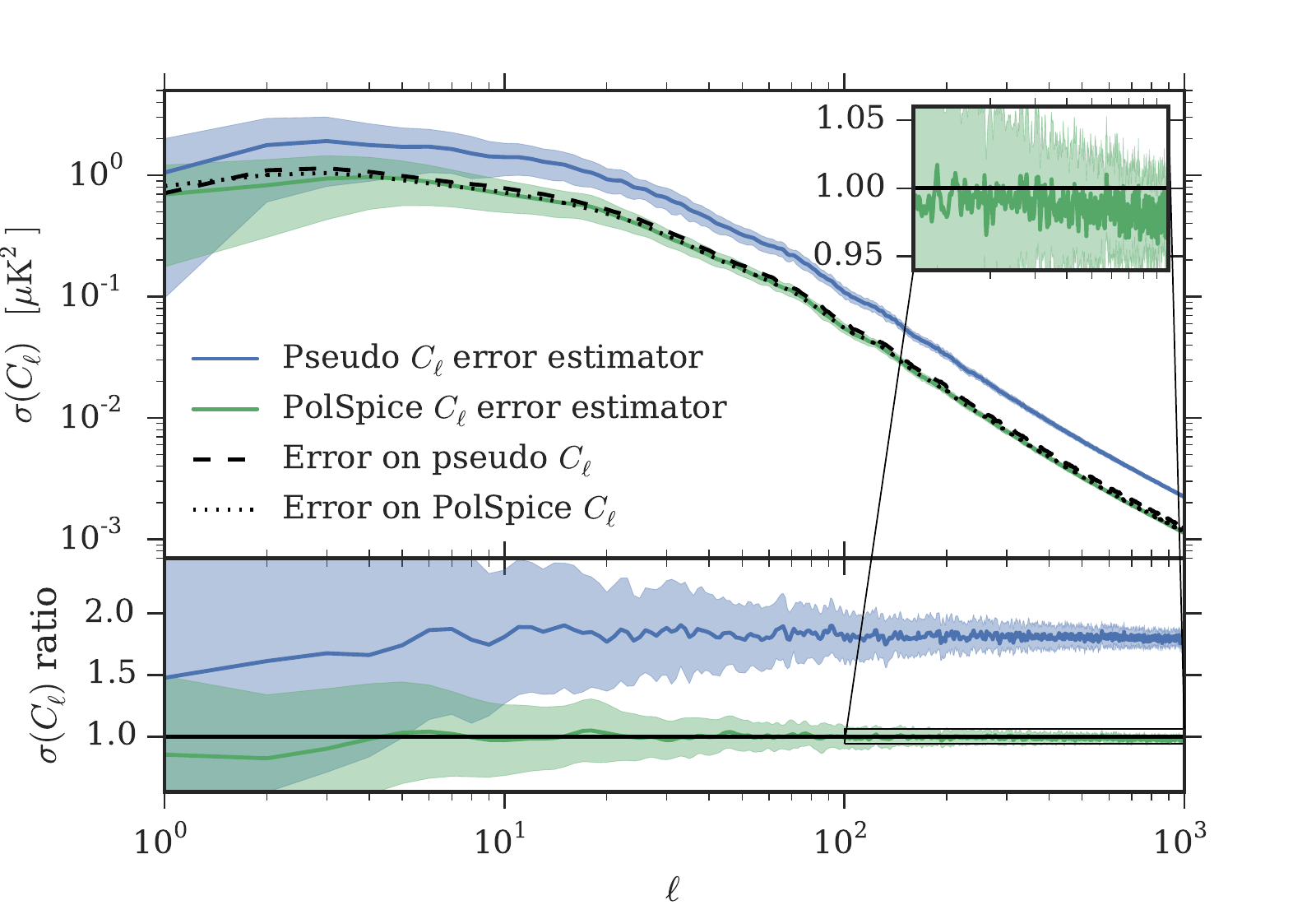}
	\caption{Standard deviation $\sigma(C_\ell)$ of the angular power spectrum from multiple realizations of the masked log-normal field as compared to predictions from Gaussian field statistics for pseudo $C_\ell$ and PolSpice estimators. While the pseudo $C_\ell$ is again biased to higher $\sigma(C_\ell)$ by a factor of $2$, the PolSpice estimator only shows a mild deviation of a view percent at small scales (inset in the top right corner).}
	\label{fig:lgn}
\end{figure}

We want to use a log-normal field with correlations that are close to the prediction from the intensity map of the lowest redshift slice ($z \approx 0.366 \pm 0.026$) in order to approximate its non-Gaussian nature. As we cannot easily simulate a log-normal field with specified power spectrum, we start from a Gaussian field created from an input power spectrum given by the linear theory prediction at this redshift. When exponentiating a Gaussian field $X$ with mean $\mu = 0$ and variance $\sigma^2$, the resulting field is log-normally distributed with mean $\exp(\sigma^2/2)$ and variance $(\exp(\sigma^2) - 1)\exp(\sigma^2)$. The following field $Y$ can hence be shown to be log-normally distributed with variance $\sigma_T^2$ and mean equal to $T$:
\begin{equation}
	Y = T \exp\left(\frac {\sqrt{\log((\sigma_T/T)^2 + 1)}} \sigma X - \frac 1 2 \log((\sigma_T/T)^2 + 1)\right),
\end{equation}
We use the map as given by $Y$ as an approximation for the desired intensity map, but need to be aware that its angular power spectrum is not necessarily consistent with the input.

\begin{figure}[t]
	\centering
	\includegraphics[width = .6\linewidth]{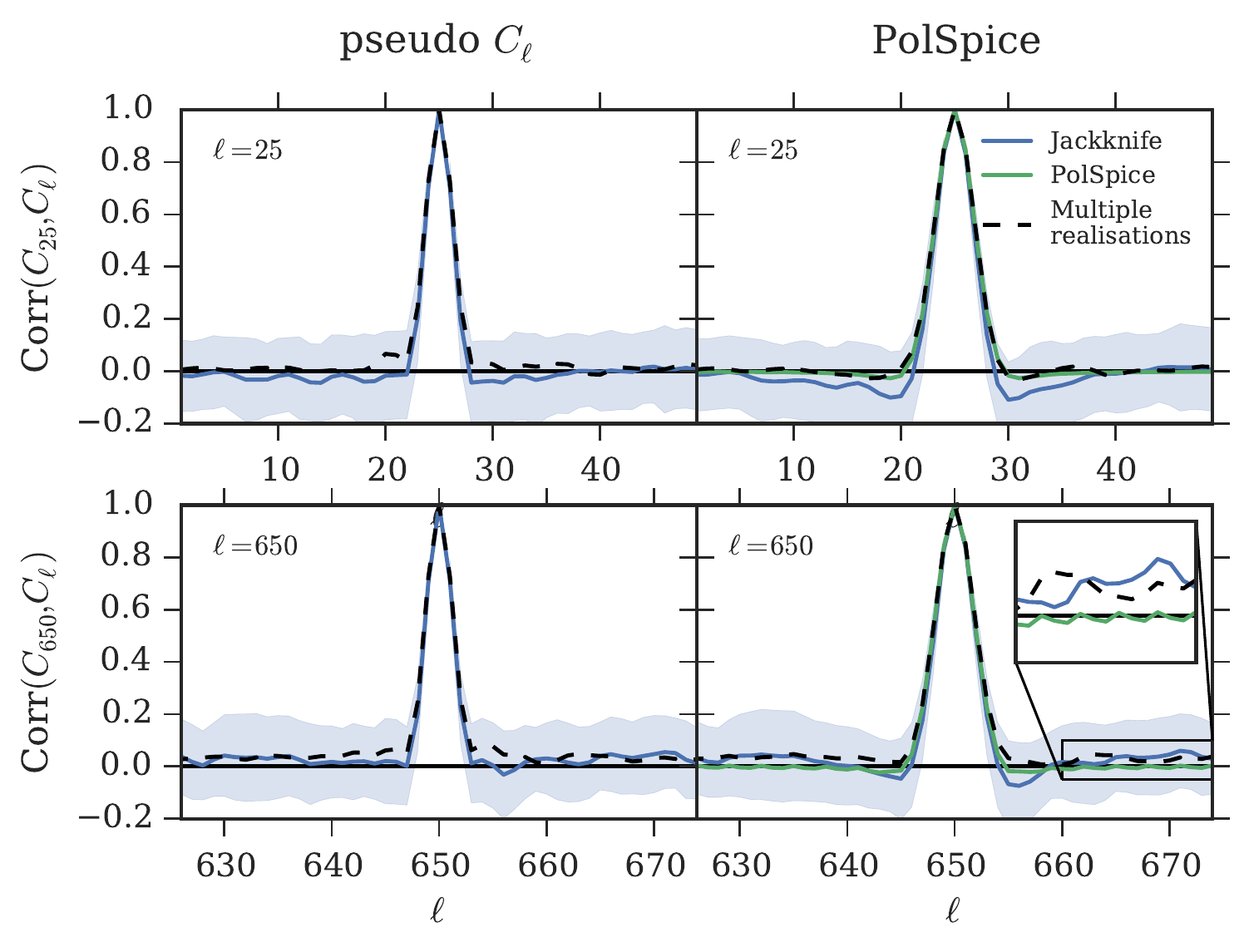}
	\caption{Correlations of the angular power spectrum from multiple realizations of the masked log-normal field as compared to estimates from jackknife and Gaussian field statistics for pseudo $C_\ell$s (left) and PolSpice (right) for large scales ($\ell = 25$, top) and small scales ($\ell = 650$, bottom).}
	\label{fig:jnlgncorrs}
\end{figure}

As shown by Figure~\ref{fig:lgn}, the variances as predicted using Gaussian field statistics for both pseudo $C_\ell$ and PolSpice are affected by the change in the underlying distribution only very mildly at small scales. As expected, the jackknife results are independent of the distribution and are equivalent to the ones shown in section \ref{sub:gaussian_fields}.

\begin{figure}[t]
	\centering
	\includegraphics[width = .8\linewidth]{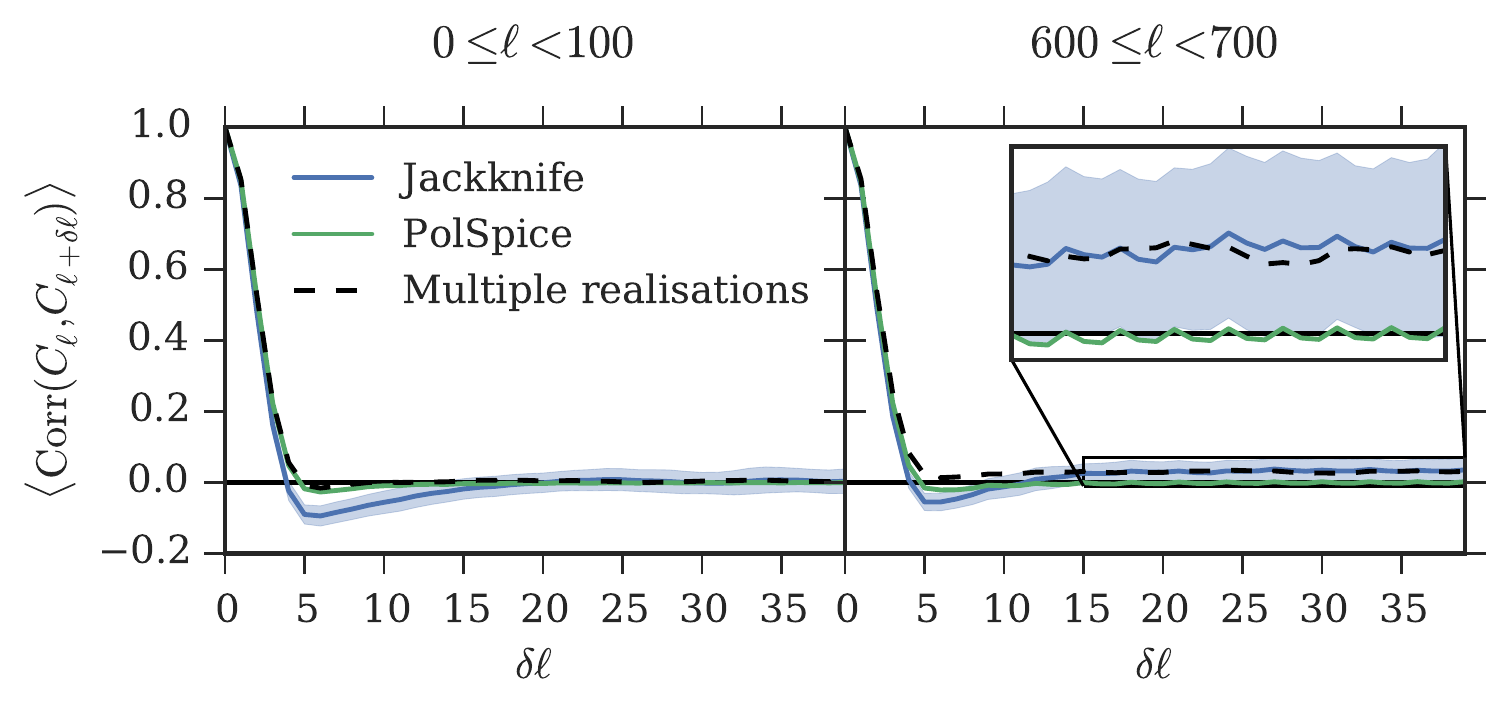}
	\caption{Mean correlation $\langle\text{Corr}(C_{\ell}, C_{\ell+\delta\ell})\rangle$ for the PolSpice angular power spectrum of the masked log-normal field, averaged over 100 $\ell$s. We show estimates from the multiple realizations, the jackknife, and the predictions for Gaussian fields. Correlations between large scales ($0 \leq \ell < 100$) are shown on the left, small scales ($600 \leq \ell < 700$) on the right.}
	\label{fig:lgnmeancorr}
\end{figure}

We now turn our attention to the correlation matrix ${\rm Corr}(C_{\ell}, C_{\ell'}) = \frac {\text{Cov}(C_{\ell}, C_{\ell'})} {\sigma(C_\ell)\sigma(C_{\ell'})}$. Figure~\ref{fig:jnlgncorrs} shows the correlations between individual $\ell$ values at large scales around $\ell = 25$ and small scales around $\ell = 650$. As expected, the large scale part of the correlation matrix is not affected by the non-Gaussianity as the perturbations on large scales are still approximately Gaussian. Looking at the correlations between individual $\ell$ values on small scales, however, we find small amounts of extra correlations introduced by the log-normal field for scales separated by $\delta \ell > 5$. Due to the assumption of an isotropic Gaussian field, the extra correlations are not estimated correctly by PolSpice. The jackknife estimator, however, is able to pick these correlations up for $\delta \ell \gtrsim 15$ while it is biased to negative correlations for $5 \lesssim \delta \ell \lesssim 15$. The variance of the jackknife estimator (shown by the shaded area) is too large to resolve the extra correlations when estimated from a single realization of the field. Figure~\ref{fig:lgnmeancorr} hence shows the correlation $\text{Corr}(C_\ell,C_{\ell + \delta \ell})$ as a function of $\delta\ell$ when averaged over 100 neighboring $\ell$ values. Again, there are additional correlations at small scales ($600 \leq \ell < 700$, right panel), but this time the noise in the jackknife estimate is just about small enough to detect those correlations (for $\delta \ell \gtrsim 15$) even from a single realization.

\end{document}